\documentclass[10pt]{amsart}

\usepackage[francais]{babel}
\usepackage[T1]{fontenc}
\usepackage[utf8]{inputenc}
\usepackage{amsmath,amssymb,mathrsfs}
\usepackage{amscd}
\usepackage{tikz}

\numberwithin{equation}{section}
\newtheorem{theo}{Théorème}[section]
\newtheorem{prop}{Proposition}[section]
\newtheorem{lem}{Lemme}[section]
\newtheorem{déf}{Définition}[section]
\newtheorem{rem}{Remarque}[section]
\numberwithin{figure}{section}

\newenvironment{preuve}
	{\textit{\textbf{Preuve.}}}
	{\hfill $\blacksquare$\vskip 8pt}

\title[Résonances près de seuils d'opérateurs de Pauli et de Dirac]{Résonances près de seuils d'opérateurs magnétiques de Pauli et de Dirac}
\author{Diomba \textsc{Sambou}}
\address{Univ. Bordeaux, Institut de Mathématiques de Bordeaux, UMR 5251, Université de Bordeaux I, F-3340 Talence, France.}
\email{diomba.sambou@math.u-bordeaux1.fr}


\begin{document}

\begin{abstract}
Nous considérons les perturbations $H := H_{0} + V$ et $D := D_{0} + V$ des Hamiltoniens libres $H_{0}$ de Pauli et $D_{0}$ de Dirac en dimension 3 avec champ magnétique non constant, $V$ étant un potentiel électrique qui décroît super-exponentiellement dans la direction du champ magnétique. Nous montrons que dans des espaces de Banach appropriés, les résolvantes de $H$ et $D$ définies sur le demi-plan supérieur admettent des prolongements méromorphes. Nous définissons les résonances de $H$ et $D$ comme étant les pôles de ces extensions méromorphes. D'une part, nous étudions la répartition des résonances de $H$ près de l'origine $0$ et d'autre part, celle des résonances de $D$ près de $\pm m$ où $m$ est la masse d'une particule. Dans les deux cas, nous obtenons d'abord des majorations du nombre de résonances dans de petits domaines au voisinage de $0$ et $\pm m$. Sous des hypothèses supplémentaires, nous obtenons des développements asymptotiques du nombre de résonances qui entraînent leur accumulation près des seuils $0$ et $\pm m$. En particulier, pour une perturbation $V$ de signe défini, nous obtenons des informations sur la répartition des valeurs propres de $H$ et $D$ près de $0$ et $\pm m$ respectivement.
\\\\
ABSTRACT. We consider the perturbations $H := H_{0} + V$ and $D := D_{0} + V$ of the free $3$D Hamiltonians $H_{0}$ of Pauli and $D_{0}$ of Dirac with non-constant magnetic field, and $V$ is a electric potential which decays super-exponentially with respect to the variable along the magnetic field. We show that in appropriate Banach spaces, the resolvents of $H$ and $D$ defined on the upper half-plane admit meromorphic extensions. We define the resonances of $H$ and $D$ as the poles of these meromorphic extensions. We study the distribution of resonances of $H$ close to the origin $0$ and that of $D$ close to $\pm m$, where $m$ is the mass of a particle. In both cases, we first obtain an upper bound of the number of resonances in small domains in a vicinity of $0$ and $\pm m$. Moreover, under additional assumptions, we establish asymptotic expansions of the number of resonances which imply their accumulation near the thresholds $0$ and $\pm m$. In particular, for a perturbation $V$ of definite sign, we obtain information on the distribution of eigenvalues of $H$ and $D$ near $0$ and $\pm m$ respectively.
\end{abstract}

\maketitle

\newpage

\section{Introduction}

Dans cet article, le but est d'étudier les résonances (ou valeurs propres) près de $0$ de l'opérateur de Pauli perturbé $H$ défini par \eqref{eq1,15}, et près de $\pm m$ de l'opérateur de Dirac perturbé $D$ défini par \eqref{eq1,16}. La perturbation $V \equiv \lbrace V_{jk} \rbrace_{1 \leq j,k \leq n}$ $(n = 2$ ou $4)$ est un potentiel matriciel hermitien symétrique identifié à l'opérateur de multiplication par $V$, et dont les coefficients $V_{jk} \in L^{\infty}(\mathbb{R}^{3},\mathbb{C})$ décroissent super-exponentiellement par rapport à la variable $x_{3}$. Pour ces opérateurs, G. D. Raikov dans \cite{raik} et R. Tiedra de Aldecoa dans \cite{tda} étudient la fonction de décalage spectral respectivement près de $0$ et $\pm m$ et montrent qu'elle possède des singularités pour $V$ de signe fixé décroissant polynomialement à l'infini. Il est naturel de penser que ces explosions de la fonction de décalage spectral sont liées à une accumulation de résonances près de $0$ et $\pm m$. Cela a d'ailleurs été montré pour l'opérateur de Schrödinger avec champ magnétique constant par J. F. Bony, V. Bruneau et G. D. Raikov dans \cite{bon} et \cite{bo}, suite à un travail de C. Fernandez et G. D. Raikov \cite{fer} sur la fonction de décalage spectral près des niveaux de Landau.

En suivant la démarche de \cite{bon}, nous montrons d'abord que l'étude des résonances près des seuils $0$ et $\pm m$ peut se ramener à l'étude des zéros d'un certain déterminant (Propositions \ref{prop2,16} et \ref{prop3,30}). Dans les Théorèmes \ref{theo2,18} et \ref{theo3,31}, nous donnons une majoration du nombre de résonances près de $0$ et près de $\pm m$ respectivement. Cette majoration est décrite en termes du nombre de valeurs propres d'un certain opérateur compact de type Toeplitz. Dans les Théorèmes \ref{theo2,24} et \ref{theo3,32} et pour $V$ de signe défini vérifiant l'hypothèse de décroissance (\ref{eq1,14}), nous donnons respectivement un développement asymptotique du nombre de résonances près de $0$ et près de $\pm m$. Nous en déduisons en particulier une accumulation des résonances près des seuils $0$ et $\pm m$.

Nous supposons que le champ magnétique $B: \mathbb{R}^{2} \longrightarrow \mathbb{R}^{3}$ est unidirectionnel de la forme $B \equiv (0,0,b)$ où la composante $b$ est dans la classe des champs électromagnétiques admissibles définie dans la Section 2.1 de \cite{raik}. En d'autres termes, nous supposons que $b = b_{0} + \tilde{b}$ où $b_{0} > 0$ est une constante et $\tilde{b} : \mathbb{R}^{2} \rightarrow \mathbb{R}$ une fonction telle que l'équation de Poisson $\Delta \tilde{\varphi} = \tilde{b}$ admette une solution $\tilde{\varphi} : \mathbb{R}^{2} \rightarrow \mathbb{R}$, continue bornée ainsi que ses dérivées d'ordre $\leq 2$. \\ 
Pour tout $x_{12} := (x_{1},x_{2}) \in \mathbb{R}^{2}$, nous définissons également $\varphi_{0} (x_{12}):= \frac{1}{4}b_{0}\vert x_{12} \vert^{2}$ et posons $\varphi := \varphi_{0} + \tilde{\varphi}$. Nous obtenons ainsi un potentiel magnétique $A \equiv (A_{1},A_{2},A_{3}) \in C^{1}(\mathbb{R}^{2},\mathbb{R}^{3})$ générant le champ magnétique $B$ en considérant $$A_{1} := -\partial_{x_{2}} \varphi, \hspace{1cm} A_{2} := \partial_{x_{1}} \varphi, \hspace{1cm} A_{3} := 0.$$
Posons
\begin{equation}\label{eq1,3}
\Pi_{1} := -i\partial_{x_{1}} - A_{1}, \hspace{1cm} \Pi_{2} := -i\partial_{x_{2}} - A_{2}, \hspace{1cm}  \Pi_{3} := -i\partial_{x_{3}}.
\end{equation}
Soient les matrices $4 \times 4$ standards de Dirac $\alpha = (\alpha_{1},\alpha_{2},\alpha_{3})$ et $\beta$. Pour $j \in \lbrace 1,2,3 \rbrace$
\begin{equation}\label{eq1,4}
\alpha_{j} := \begin{pmatrix}
   \textbf{0} & \sigma _{j}\\
   \sigma_{j} & \textbf{0}
\end{pmatrix}, \quad \beta := \begin{pmatrix}
   \textbf{1} & \textbf{0} \\
   \textbf{0} & \textbf{-1}
\end{pmatrix},
\end{equation}
où $\textbf{0}$ et $\textbf{1}$ sont les matrices $2 \times 2$ nulle et unité, les $\sigma_{j}$ sont les matrices $2 \times 2$ de Pauli définies par
\begin{equation}\label{eq1,5}
\sigma_{1} = \begin{pmatrix}
   0 & 1 \\
   1 & 0
\end{pmatrix}, \hspace{0.5cm} \sigma_{2} = \begin{pmatrix}
   0 & -i \\
   i & 0
\end{pmatrix}, \hspace{0.5cm} \sigma_{3} = \begin{pmatrix}
   1 & 0 \\
   0 & -1
\end{pmatrix}.
\end{equation}
Notons que le choix des matrices $\alpha$ et $\beta$ n'est pas unique (voir p. e. l'appendice du Chapitre 1 de \cite{tha} pour d'autres représentations). Les matrices $\alpha_{j}$ et $\beta$ sont en fait déterminées par les relations suivantes pour $j, k \in \lbrace 1,2,3 \rbrace$:
\begin{equation}\label{eq1,6}
\alpha_{j}\alpha_{k} + \alpha_{k}\alpha_{j} = 2\delta_{jk}\textbf{1}, \hspace{1cm} \alpha_{j}\beta + \beta\alpha_{j} = \textbf{0}, \hspace{1cm} \beta^{2}= \textbf{1},
\end{equation}
où $\delta_{jk}$ est le symbole de Kronecker défini par $\delta_{jk} = 1$ si $j = k$ et $\delta_{jk} = 0$ si $j \neq k$. 

Pour $n = 2$ ou $4$, soit $L^{2}(\mathbb{R}^{3}) := L^{2}(\mathbb{R}^{3},\mathbb{C}^{n}) = L^{2}(\mathbb{R}^{2},\mathbb{C}^{n}) \otimes L^{2}(\mathbb{R})$. Désignons par $\textup{a}$ et $\textup{a}^{\ast}$ les fermetures dans $L^{2}(\mathbb{R}^{2}) := L^{2}(\mathbb{R}^{2},\mathbb{C})$ des opérateurs définis sur $C_{0}^{\infty}(\mathbb{R}^{2})$ par 
\begin{equation}\label{eq1,7}
\textup{a} := \Pi_{1} + i\Pi_{2} = -2i \textup{e}^{-\varphi} \partial_{\bar{z}} \textup{e}^{\varphi}, \hspace{1cm} \textup{a}^{\ast} := \Pi_{1} - i\Pi_{2} = -2i \textup{e}^{\varphi} \partial_{z} \textup{e}^{-\varphi},
\end{equation}
où les $\Pi_{j}$ sont définis par \eqref{eq1,3}, $z = x_{1} + i x_{2}$ et $\bar{z} = x_{1} - i x_{2}$. Posons 
\begin{equation}\label{eq1,8}
H_{12}^{+} : = \textup{a} \textup{a}^{\ast}, \hspace{1cm} H_{12}^{-} : = \textup{a}^{\ast} \textup{a}.
\end{equation}
On a (voir sous-section 2.2 de \cite{raik})
\begin{equation}\label{eq1.161}
\begin{split}
\textup{ker} \hspace{0.5mm} H_{12}^{-} & = \textup{ker}(\textup{a}) = \left\lbrace u \in L^{2}(\mathbb{R}^{2}) : u = g \textup{e}^{-\varphi} \hspace{1mm} et \hspace{1mm} \partial_{\bar{z}} g = 0 \right\rbrace,\\
\textup{ker} \hspace{0.5mm} H_{12}^{+} & = \textup{ker}(\textup{a}^{\ast}) = \left\lbrace u \in L^{2}(\mathbb{R}^{2}) : u = g \textup{e}^{\varphi} \hspace{1mm} et \hspace{1mm} \partial_{z} g = 0 \right\rbrace,
\end{split}
\end{equation}
$\dim \hspace{0.5mm} \textup{ker} \hspace{0.5mm} H_{12}^{-} = \infty$, $\dim \hspace{0.5mm} \textup{ker} \hspace{0.5mm} H_{12}^{+} = 0$, $\sigma (H_{12}^{\pm}) \subset \lbrace 0 \rbrace \cup [\zeta,\infty)$ où 
\begin{equation}\label{eq0}
\zeta := 2 b_{0} \textup{exp} (-2 \textup{osc} \hspace{0.5mm} \tilde{\varphi}) > 0
\end{equation}
et 
\begin{equation}\label{eq00}
\textup{osc} \hspace{0.5mm} \tilde{\varphi}:= \displaystyle\sup_{x_{12} \in \mathbb{R}^{2}} \tilde{\varphi} (x_{12}) - \displaystyle\inf_{x_{12} \in \mathbb{R}^{2}} \tilde{\varphi} (x_{12}).
\end{equation}
Par ailleurs, il est bien connu que la projection orthogonale $p$ sur $\textup{ker} \hspace{0.5mm} H_{12}^{-}$ admet un noyau intégral continu $\mathcal{P}(x_{12},y_{12})$ où $x_{12}$, $y_{12} \in \mathbb{R}^{2}$ (Théorème 2.3 de \cite{hal}).

L'opérateur $H_{0}$ de Pauli est défini a priori sur $C_{0}^{\infty}(\mathbb{R}^{3},\mathbb{C}^{2})$ par 
\begin{equation}\label{eq1,9}
H_{0} := (\sigma \cdot (-i\nabla - A))^{2},
\end{equation}
où $\sigma = (\sigma_{1},\sigma_{2},\sigma_{3})$ est le triplet de matrices de Pauli définies par \eqref{eq1,5}. Il y est essentiellement auto-adjoint de spectre $\sigma(H_{0}) = [0,\infty)$ (voir \cite{raik}). D'autres résultats sur le spectre d'opérateurs de Pauli peuvent être trouvés dans \cite{los} et \cite{ra}. Une représentation matricielle de $H_{0}$ est donnée par
\begin{equation}\label{eq1,10}
H_{0} = \begin{pmatrix}
   H_{12}^{-} \otimes 1 + 1 \otimes \Pi_{3}^{2} & 0\\
   0 & H_{12}^{+} \otimes 1 + 1 \otimes \Pi_{3}^{2}
\end{pmatrix} =: \begin{pmatrix}
   H_{0}^{-} & 0 \\
   0 & H_{0}^{+}
\end{pmatrix}.
\end{equation}

L'opérateur $D_{0}$ de Dirac est défini a priori sur $C_{0}^{\infty}(\mathbb{R}^{3},\mathbb{C}^{4})$ par 
\begin{equation}\label{eq1,11}
\begin{aligned}
D_{0} & := \alpha \cdot (-i\nabla - A) + m \beta \\
& = \alpha_{1} \Pi_{1} + \alpha_{2} \Pi_{2} + \alpha_{3} \Pi_{3} + m \beta,
\end{aligned}
\end{equation}
où les $\alpha_{j}$ et $\beta$ sont les matrices de Dirac définies par \eqref{eq1,4}, les $\Pi_{j}$ sont définies par \eqref{eq1,3}, $m > 0$ est la masse d'une particule. Il y est essentiellement auto-adjoint et nous avons $\sigma (D_{0}) = (-\infty,-m] \cup [m,\infty)$ (voir \cite{tda}). Dans \cite{geo}, \cite{hel}, \cite{ric}, \cite{tha}, on peut trouver d'autres résultats sur le spectre d'opérateurs de Dirac et de plus dans \cite{ada}, \cite{mur}, \cite{nas}, \cite{sa}, \cite{ume}, \cite{sai} des résultats sur les zéro-modes. Une expression explicite matricielle de $D_{0}$ est donnée par 
\begin{equation}\label{eq1,12}
D_{0} = \begin{pmatrix}
   m & 0 & 1 \otimes \Pi_{3} & \textup{a}^{\ast} \otimes 1\\
   0 & m & \textup{a} \otimes 1 & -1 \otimes \Pi_{3}\\
   1 \otimes \Pi_{3} & \textup{a}^{\ast} \otimes 1 & -m & 0\\
   \textup{a} \otimes 1 & -1 \otimes \Pi_{3} & 0 & -m
\end{pmatrix}
\end{equation}
et nous avons l'identité
\begin{equation}\label{eq1,13}
D_{0}^{2} = \left( \begin{smallmatrix}
H_{12}^{-} \otimes 1 + 1 \otimes (\Pi_{3}^{2} + m^{2}) & 0 & 0 & 0\\
   0 & H_{12}^{+} \otimes 1 + 1 \otimes (\Pi_{3}^{2} + m^{2}) & 0 & 0\\
   0 & 0 & H_{12}^{-} \otimes 1 + 1 \otimes (\Pi_{3}^{2} + m^{2}) & 0\\
   0 & 0 & 0 & H_{12}^{+} \otimes 1 + 1 \otimes (\Pi_{3}^{2} + m^{2})
\end{smallmatrix} \right).
\end{equation}

Dans la suite, pour un espace de Hilbert $X$ séparable et $q \in [1,\infty]$, $S_{q}(X)$ désigne la classe de Schatten-Von Neumann des opérateurs compacts $T$ sur $X$ pour lesquels la norme $\Vert T \Vert_{q} := (\textup{Tr} \hspace{0.5mm} \vert T \vert^{q})^{1/q}$ est finie ($S_{\infty}(X)$ est l'ensemble des opérateurs compacts sur $X$). Pour $q = 1$, $S_{1}(X)$ est l'espace de Banach des opérateurs à trace; pour $q = 2$, $S_{2}(X)$ est l'espace de Banach des opérateurs de Hilbert-Schmidt. 

Nous adoptons également le choix standard de la racine carrée complexe\\ $\sqrt{\cdot} : \mathbb{C} \setminus [0,+\infty) \longrightarrow \mathbb{C}^{+} := \lbrace \zeta \in \mathbb{C} \setminus \textup{Im} \hspace{0.5mm} \zeta > 0 \rbrace$ dans tout l'article.

Le plan adopté est le suivant. Nos résultats principaux sont présentés dans la Section 2. Dans la Section 3, nous rappelons quelques résultats auxiliaires sur les opérateurs de type Toeplitz et sur les valeurs caractéristiques d'une fonction holomorphe à valeur opérateur. Dans la Section 4, nous définissons les résonances de l'opérateur $H$ près de $0$ et prouvons les Théorèmes \ref{theo2,18} et \ref{theo2,24}. Dans la Section 5, sont définies les résonances de $D$ près de $\pm m$. Par ailleurs, nous y prouvons les Théorèmes \ref{theo3,31} et \ref{theo3,32}. La Section 6 est un bref appendice sur les notions d'indice (le long d'un contour fermé orienté positivement) d'une fonction holomorphe et d'une fonction méromorphe finie à valeur opérateur.

\section{Énoncés des résultats principaux}

Pour tout $x = (x_{12},x_{3}) \in \mathbb{R}^{3}$, nous supposons que $V \equiv \lbrace V_{jk} \rbrace_{1\leq j,k \leq n}$ satisfait
\begin{equation}\label{eq1,14}
V(x) \in \mathfrak{B}_{\textup{h}}(\mathbb{C}^{n}), \hspace{0.5cm} \vert V_{jk}(x) \vert = O \left( \langle x_{12} \rangle^{-m_{12}} \hspace{0.5mm} \textup{exp} ( -2 \delta \langle x_{3} \rangle ) \right),
\end{equation} 
où $m_{12} > 0$, $\delta > 0$, $\langle x_{12} \rangle := \left(1 + \vert x_{12} \vert^{2}\right)^{1/2}$ et $\mathfrak{B}_{\textup{h}}(\mathbb{C}^{n})$ est l'ensemble des matrices $n \times n$ hermitiennes symétriques. 

Pour $n = 2$, nous définissons dans le domaine de $H_{0}$ l'opérateur
\begin{equation}\label{eq1,15}
H := H_{0} + V.
\end{equation}
$\hspace{0.35cm}$ Pour $n = 4$, nous définissons dans le domaine de $D_{0}$ l'opérateur
\begin{equation}\label{eq1,16}
D := D_{0} + V.
\end{equation}

Soient les opérateurs $p W_{\pm} p$ où $p$ est la projection orthogonale sur $\textup{ker} \hspace{0.5mm} H_{12}^{-}$ défini par \eqref{eq1.161}, et $W_{\pm}$ sont les opérateurs de multiplication par les fonctions $W_{\pm} : \mathbb{R}^{2} \rightarrow \mathbb{R}$ définies par 
\begin{equation}\label{eq2,22}
\begin{split}
 W_{+}(x_{12}) := \int_{\mathbb{R}} \vert V_{11} \vert (x_{12},& x_{3}) dx_{3}  \quad \textup{et} \quad W_{-}(x_{12}) := \int_{\mathbb{R}} \vert V_{33} \vert (x_{12},x_{3}) dx_{3},\\ 
& W_{+} =: W \quad \text{pour} \quad n = 2,
\end{split}
\end{equation}
où les $\vert V_{jk} \vert$ sont par définition les coefficients de la matrice $\vert V \vert$. Sous l'hypothèse \textup{(\ref{eq1,14})}, pour tout $x_{12} \in \mathbb{R}^{2}$, $0 \leq \vert W_{\pm}(x_{12}) \vert = O \left(\langle x_{12} \rangle^{-m_{12}} \right)$. Le Lemme \ref{lem1,1} ci-dessous entraîne ainsi que les opérateurs $p W_{\pm} p$ sont compacts dans $L^{2}(\mathbb{R}^{2})$.

Pour un opérateur auto-adjoint $A$ compact, posons pour tout $s > 0$
\begin{equation}\label{eq2,222}
n_{+}(s,A) := \textup{rang} \hspace{0.5mm} \textup{\textbf{1}}_{(s,+\infty)}(A).
\end{equation}
Soit le disque pointé
\begin{equation}\label{eq2,110}
D(0,\epsilon)^{\ast} := \lbrace k \in \mathbb{C} : 0 < \vert k \vert < \epsilon \rbrace
\end{equation}
avec
$$\epsilon < \min \hspace{0.5mm} \left( \delta,\sqrt{\zeta} \right),$$
où $\delta$ et $\zeta$ sont respectivement définis par \eqref{eq1,14} et \eqref{eq0}. Désignons par $\textup{Res}(H)$ l'ensemble des résonances de $H$ près de $0$ (Définition \ref{def2,11}). Notre premier résultat est une majoration du nombre de résonances $z(k) = k^{2}$ de $H$ près de $0$ pour $k \in D(0,\epsilon)^{\ast}$.

\begin{theo}\label{theo2,18} 
$\textup{(Borne supérieure)}$ Supposons que \eqref{eq1,14} soit vérifiée pour $n = 2$ et soit $W$ défini par \eqref{eq2,22}. Alors il existe $r_{0} > 0$ tel que pour tout $0 < r < r_{0}$, 
\begin{equation}\label{eq2,181}
\# \lbrace z = z(k) := k^{2} \in \textup{Res}(H) : r < \vert k \vert < 2r \rbrace = O \big{(} n_{+}(r,pWp) \vert \textup{ln} \hspace{0.5mm} r \vert \big{)} + O(1),
\end{equation}
où si la fonction $W$ satisfait les hypothèses des \textup{Lemmes \ref{lem2,13}, \ref{lem2,14} et \ref{lem2,15}} nous avons respectivement:
$$\textup{(i)} \hspace{0.5cm} n_{+}(r,pWp) = O \big{(} r^{-2/m_{12}} \big{)},$$

$$\textup{(ii)} \hspace{0.5cm} n_{+} (r,p W p) = O \big{(} \varphi_{\beta}(r) \big{)}$$ avec $$\small{\varphi_{\beta}(r) :=
 \begin{cases}
 \frac{1}{2} b_{0} \mu^{-1/\beta} \vert \ln r \vert^{1/\beta} & \text{si } 0 < \beta < 1,\\
 \dfrac{1}{\ln(1 + 2\mu/b_{0})} \vert \ln r \vert & \text{si } \beta = 1,\\
 \dfrac{\beta}{\beta - 1} \big{(} \ln \vert \ln r \vert \big{)}^{-1} \vert \ln r \vert & \text{si } \beta > 1.
 \end{cases}}, \hspace*{0.5cm} \mu > 0,$$

$$\textup{(iii)} \hspace{0.5cm} n_{+}(r,pWp) = O \big{(} (\ln \vert \ln r \vert)^{-1} \vert \ln r \vert \big{)}.$$
\end{theo}

Notre second résultat concerne un développement asymptotique près de $0$ du nombre de résonances de l'Hamiltonien perturbé $H_{e} := H_{0} + eV$ ($V > 0$), où $e \in \mathbb{R}^{\ast} \setminus E$ avec $E$ un ensemble discret de $\mathbb{R}^{\ast}$.

\begin{theo}\label{theo2,24} 
$\textup{(Développement asymptotique)}$ Supposons que \eqref{eq1,14} soit vérifiée pour $n = 2$, $V > 0$ et soit $W$ défini par \eqref{eq2,22}. Il existe un ensemble discret $E$ de $\mathbb{R}^{\ast}$ tel que pour tout $e \in \mathbb{R}^{\ast} \setminus E$, $H_{e} := H_{0} + eV$ possède les propriétés suivantes:

$\textup{(i)}$ Près de $0$, les résonances $z(k) := k^{2}$ de $H_{e}$ pour $\vert k \vert < \epsilon$ assez petit vérifient $$e \hspace{0.5mm} \textup{Im} \hspace{0.5mm} k \leq 0, \hspace{0.7cm} \textup{Re} \hspace{0.5mm} k = o(\vert k \vert).$$

$\textup{(ii)}$ Il existe une suite $(r_{l})_{l} \in \mathbb{R}^{\mathbb{N}}$ tendant vers $0$ telle que  $$\# \lbrace z = z(k) \in \textup{Res}(H_{e}) : r_{l} < \vert k \vert \leq r_{0} \rbrace = n_{+} \left( r_{l},\frac{1}{2}pWp \right) (1 + o(1)).$$ 

$\textup{(iii)}$ Si $W = U$ satisfait les hypothèses du \textup{Lemme} $\ref{lem2,13}$ ou $\ref{lem2,14}$ ou $\ref{lem2,15}$, $$\# \lbrace z = z(k) \in \textup{Res}(H_{e}) : r < \vert k \vert \leq r_{0} \rbrace = n_{+} \left( r,\frac{1}{2}pWp \right) (1 + o(1)), \hspace{0.2cm} r \searrow 0.$$
\end{theo}

\begin{rem}\label{rem2,1} 
\textup{Plus généralement dans le Théorème \ref{theo2,24}, l'hypothèse $V > 0$ peut être remplacée par l'hypothèse $(\textup{sign} \hspace{0.5mm} V) \tiny{\begin{pmatrix}
1 & 0\\
0 & 0
\end{pmatrix}} = \tiny{\begin{pmatrix}
1 & 0\\
0 & 0
\end{pmatrix}}$. Par exemple si le potentiel $V = \textup{Diag} \hspace{0.5mm} (V_{11},V_{22})$ est diagonal et $V_{11} > 0$, alors cette dernière hypothèse est satisfaite.}
\end{rem}
Près de $0$, les valeurs propres de l'opérateur $H$ sont les résonances $z(k)$ avec $k \in \textup{e}^{i\lbrace 0,\frac{\pi}{2} \rbrace} ]0,\epsilon[$. Le Théorème \ref{theo2,18} fournit donc une majoration du nombre de valeurs propres de $H$ dans des intervalles du type $[-4r^{2},-r^{2}]$ pour tout $0 < r < r_{0} < \epsilon$.

Pour tout $e \in \mathbb{R}^{\ast} \setminus E$, le Théorème \ref{theo2,24} montre en particulier qu'il y a une accumulation de résonances de $H_{e} := H_{0} + eV$ près de l'origine $0$. Pour $V < 0$ ($V_{11} < 0$ suffit si $V$ est diagonale par la \textit{Remarque} \ref{rem2,1}), les seules résonances sont les valeurs propres $z(k)$ avec $k \in \textup{e}^{i\frac{\pi}{2}} ]0,\epsilon[$ et $\vert k \vert$ assez petit. Pour $V > 0$ ($V_{11} > 0$ suffit si $V$ est diagonale), $H_{e}$ n'a pas de spectre discret négatif. Nous pouvons comparer nos résultats à ceux de \cite{raik} sur la distribution asymptotique près de $0$ du spectre discret (négatif) de $H_{-} := H_{0} - V$ avec $V \geq 0$. En effet par un développement asymptotique près de $0$ de la fonction de décalage spectrale associée à la paire d'opérateurs $(H_{-},H_{0})$, dans le Corollaire 3.6 de \cite{raik} G. D. Raikov montre en particulier la même accumulation de valeurs propres de l'opérateur $H_{-}$ près de $0$ si les coefficients $V_{jk}(x)$, $1 \leq j,k \leq 2$ de la perturbation $V$ sont de l'ordre de $O \left( \langle x \rangle^{-m} \right)$ avec $m > 3$.

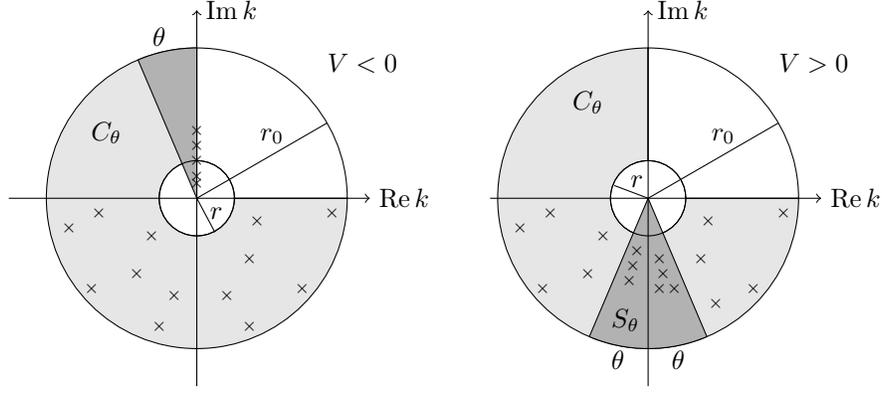
\begin{figure}\label{fig 1}
\begin{center}
\tikzstyle{+grisEncadre}=[fill=gray!60]
\tikzstyle{blancEncadre}=[fill=white!100]
\tikzstyle{grisEncadre}=[fill=gray!20]
\begin{tikzpicture}[scale=1]
\begin{scope}
\draw [grisEncadre] (0,0) -- (90:2) arc (90:360:2) -- cycle;
\draw [blancEncadre] (0,0) -- (0:2) arc (0:90:2) -- cycle;
\draw [blancEncadre] (0,0) circle(0.5);
\draw [+grisEncadre] (0,0) -- (90:2) arc (90:113:2) -- cycle;
\draw (0,0) circle(0.5);
\draw[->] (-2.5,0) -- (2.3,0);
\draw (2.3,0) node[right] {$\textup{Re} \hspace{0.6mm} k$};
\draw[->] (0,-2.5) -- (0,2.5);
\draw (0,2.5) node[right] {$\textup{Im} \hspace{0.6mm} k$};
\draw (0.26,-0.40) node[above] {$\tiny{r}$};
\draw (0,0) -- (0.24,-0.45);
\draw (0,0) -- (1.73,1);
\draw (1,0.6) node[above] {$r_{0}$};
\draw (-0.5,1.9) node[above] {$\theta$};
\draw (-1.2,0.6) node[above] {$C_{\theta}$};
\node at (0,0.9) {\tiny{$\times$}};
\node at (0,0.7) {\tiny{$\times$}};
\node at (0,0.5) {\tiny{$\times$}};
\node at (0,0.3) {\tiny{$\times$}};
\node at (0,0.2) {\tiny{$\times$}};
\node at (-1.7,-0.4) {\tiny{$\times$}};
\node at (-1.3,-0.2) {\tiny{$\times$}};
\node at (-1.4,-1.2) {\tiny{$\times$}};
\node at (-0.8,-1) {\tiny{$\times$}};
\node at (-0.6,-0.5) {\tiny{$\times$}};
\node at (-0.5,-1.7) {\tiny{$\times$}};
\node at (-0.3,-1.3) {\tiny{$\times$}};
\node at (1.8,-0.2) {\tiny{$\times$}};
\node at (1.4,-1.2) {\tiny{$\times$}};
\node at (0.7,-0.8) {\tiny{$\times$}};
\node at (0.8,-0.3) {\tiny{$\times$}};
\node at (0.7,-1.7) {\tiny{$\times$}};
\node at (0.4,-1.3) {\tiny{$\times$}};
\node at (2.2,1.8) {$V < 0$};
\end{scope}

\begin{scope}[xshift=6cm]
\draw [grisEncadre] (0,0) -- (90:2) arc (90:360:2) -- cycle;
\draw [blancEncadre] (0,0) -- (0:2) arc (0:90:2) -- cycle;
\draw [blancEncadre] (0,0) circle(0.5);
\draw [+grisEncadre] (0,0) -- (-113:2) arc (-113:-67:2) -- cycle;
\draw (0,0) circle(0.5);
\draw[->] (-2.1,0) -- (2.3,0);
\draw (2.3,0) node[right] {$\textup{Re} \hspace{0.6mm} k$};
\draw[->] (0,-2.5) -- (0,2.5);
\draw (0,2.5) node[right] {$\textup{Im} \hspace{0.6mm} k$};
\draw (-0.15,0.02) node[above] {$\tiny{r}$};
\draw (0,0) -- (-0.45,0.17);
\draw (0,0) -- (1.73,1);
\draw (1,0.6) node[above] {$r_{0}$};
\draw (-0.8,1) node[above] {$C_{\theta}$};
\draw (-0.3,-1.9) node[above] {$S_{\theta}$};
\draw (-0.4,-2.4) node[above] {$\theta$};
\draw (0.4,-2.4) node[above] {$\theta$};
\node at (-1.7,-0.4) {\tiny{$\times$}};
\node at (-1.3,-0.2) {\tiny{$\times$}};
\node at (-1.4,-1.2) {\tiny{$\times$}};
\node at (-0.8,-1) {\tiny{$\times$}};
\node at (-0.6,-0.5) {\tiny{$\times$}};
\node at (0.9,-1.4) {\tiny{$\times$}};
\node at (1.8,-0.2) {\tiny{$\times$}};
\node at (1.4,-1.2) {\tiny{$\times$}};
\node at (0.7,-0.8) {\tiny{$\times$}};
\node at (0.8,-0.3) {\tiny{$\times$}};
\node at (-0.15,-0.7) {\tiny{$\times$}};
\node at (-0.2,-0.9) {\tiny{$\times$}};
\node at (-0.25,-1.1) {\tiny{$\times$}};
\node at (0.15,-0.8) {\tiny{$\times$}};
\node at (0.2,-1) {\tiny{$\times$}};
\node at (0.15,-1.2) {\tiny{$\times$}};
\node at (0.35,-1.2) {\tiny{$\times$}};
\node at (2.2,1.8) {$V > 0$};
\end{scope}
\end{tikzpicture}
\caption{\textit{Localisation des résonances en variable $k$ pour $V$ de signe défini: Pour $r_{0}$ assez petit, les résonances $z(k) = k^{2}$ de $H_{0} + V$ et $z_{m}(k) = \frac{m(1+k^{2})}{1-k^{2}}$ de $D_{0} + V$ sont concentrées dans $S_{\theta}$ $\textup{(Théorème \ref{theo2,24} et \ref{theo3,32})}$. Pour $V < 0$, les seules résonances sont les valeurs propres $z(k) = k^{2}$ et $z = z_{m}(k)$ et sont concentrées sur la demi-droite $k = i]0,+\infty)$. Pour $V > 0$, elles sont concentrées près de la demi-droite $k = -i]0,+\infty)$. Dans $C_{\theta}$, le nombre de résonances est majoré par $O(\vert \ln r \vert)$ pour $V$ de signe non fixé $\textup{(Théorème \ref{theo2,18} et \ref{theo3,31})}$.}}
\end{center}
\end{figure}

Nous obtenons des résultats similaires sur les résonances de $D$ près de $\pm m$ (voir Définition \ref{def3,28}). Soient $\textup{Res}(D)$ l'ensemble des résonances $z_{\pm m}(k) := \frac{\pm m(1+k^{2})}{1-k^{2}}$ de $D$ près de $\pm m$ pour $k$ dans
\begin{equation}\label{eq2,1100}
D(0,\eta)^{\ast} := \lbrace k \in \mathbb{C} : 0 < \vert k \vert < \eta \rbrace.
\end{equation}
Le rayon $\eta$ est assez petit et vérifie
$$\eta < \min \hspace{0.5mm} \left( \frac{\delta}{4m}, \sqrt{1 - 2m/\mu} \right),$$
où $\delta$ est défini par \eqref{eq1,14} et $\mu$ est tel que $2m < \mu < r_{m} := \sqrt{m^{2} + \zeta} + m$, $\zeta$ étant défini par \eqref{eq0}.

\begin{theo}\label{theo3,31}
$\textup{(Borne supérieure)}$ Supposons que \eqref{eq1,14} soit vérifiée pour $n = 4$ et soient $W_{\pm}$ définies par \eqref{eq2,22}. Alors il existe $r_{0} > 0$ tel que pour tout $0 < r < r_{0}$, 
\begin{align*}
& \# \left\lbrace z = z_{\pm m}(k) := \frac{\pm m(1+k^{2})}{1-k^{2}} \in \textup{Res}(D) : r < \vert k \vert < 2r \right\rbrace\\
& = O \big{(} n_{+}(r,pW_{\pm }p) \vert \textup{ln} \hspace{0.5mm} r \vert \big{)} + O(1),
\end{align*}
où si les fonctions $W_{\pm}$ satisfont les hypothèses des \textup{Lemmes \ref{lem2,13}, \ref{lem2,14} et \ref{lem2,15}} nous avons respectivement:
$$\textup{(i)} \hspace{0.5cm} n_{+}(r,p W_{\pm} p) = O \big{(} r^{-2/m_{12}} \big{)},$$

$$\textup{(ii)} \hspace{0.5cm} n_{+} (r,p W_{\pm} p) = O \big{(} \varphi_{\beta}(r)$$ avec $$\small{\varphi_{\beta}(r) :=
 \begin{cases}
 \frac{1}{2} b_{0} \mu^{-1/\beta} \vert \ln r \vert^{1/\beta} & \text{si } 0 < \beta < 1,\\
 \dfrac{1}{\ln(1 + 2\mu/b_{0})} \vert \ln r \vert & \text{si } \beta = 1,\\
 \dfrac{\beta}{\beta - 1} \big{(} \ln \vert \ln r \vert \big{)}^{-1} \vert \ln r \vert & \text{si } \beta > 1.
 \end{cases}}, \hspace*{0.5cm} \mu > 0,$$

$$\textup{(iii)} \hspace{0.5cm} n_{+}(r,p W_{\pm} p) = O \big{(} (\ln \vert \ln r \vert)^{-1} \vert \ln r \vert \big{)}$$
\end{theo}

\begin{theo}\label{theo3,32} 
$\textup{(Développement asymptotique)}$ Supposons que \eqref{eq1,14} soit vérifiée pour $n = 4$, $V > 0$ et soient $W_{\pm}$ définies par \eqref{eq2,22}. Il existe un ensemble discret $\mathcal{E}$ de $\mathbb{R}^{\ast}$ tel que pour tout $\varepsilon \in \mathbb{R}^{\ast} \setminus \mathcal{E}$, $D_{\varepsilon} := D_{0} + \varepsilon V$ possède les propriétés suivantes:

$\textup{(i)}$ Près de $0$, les résonances $z_{\pm m}(k) := \frac{\pm m(1+k^{2})}{1-k^{2}}$ de $D_{\varepsilon}$ pour $\vert k \vert < \eta$ assez petit vérifient $$\pm \varepsilon \hspace{0.5mm} \textup{Im} \hspace{0.5mm} k \leq 0, \hspace{0.7cm} \textup{Re} \hspace{0.5mm} k = o(\vert k \vert).$$

$\textup{(ii)}$ Il existe une suite $(r_{l})_{l} \in \mathbb{R}^{\mathbb{N}}$ tendant vers $0$ telle que  $$\# \lbrace z = z_{\pm m}(k) \in \textup{Res}(D_{\varepsilon}) : r_{l} < \vert k \vert \leq r_{0} \rbrace = n_{+} \left( r_{l},\frac{1}{2}pW_{\pm}p \right) (1 + o(1)).$$ 

$\textup{(iii)}$ Si $W_{\pm} = U$ satisfait les hypothèses du \textup{Lemme} $\ref{lem2,13}$ ou $\ref{lem2,14}$ ou $\ref{lem2,15}$, $$\# \lbrace z = z_{\pm m}(k) \in \textup{Res}(D_{\varepsilon}) : r < \vert k \vert \leq r_{0} \rbrace = n_{+} \left( r,\frac{1}{2}pW_{\pm }p \right) (1 + o(1)), \hspace{0.2cm} r \searrow 0.$$
\end{theo}

\begin{rem}\label{rem3,33} 
\textup{De manière plus générale dans le Théorème \ref{theo3,32}, l'hypothèse $V > 0$ peut être remplacée par l'hypothèse $(\textup{sign} \hspace{0.5mm} V) \left( \begin{smallmatrix}
   1 & 0 & 0 & 0\\
   0 & 0 & 0 & 0\\
   0 & 0 & 0 & 0\\
   0 & 0 & 0 & 0
\end{smallmatrix} \right) = \left( \begin{smallmatrix}
   1 & 0 & 0 & 0\\
   0 & 0 & 0 & 0\\
   0 & 0 & 0 & 0\\
   0 & 0 & 0 & 0
\end{smallmatrix} \right)$ près de $m$. Près de $-m$, elle peut être remplacée par $(\textup{sign} \hspace{0.5mm} V) \left( \begin{smallmatrix}
   0 & 0 & 0 & 0\\
   0 & 0 & 0 & 0\\
   0 & 0 & 1 & 0\\
   0 & 0 & 0 & 0
\end{smallmatrix} \right) = \left( \begin{smallmatrix}
   0 & 0 & 0 & 0\\
   0 & 0 & 0 & 0\\
   0 & 0 & 1 & 0\\
   0 & 0 & 0 & 0
\end{smallmatrix} \right)$. Par exemple si $V = \textup{Diag} \hspace{0.5mm} (V_{11},V_{22},V_{33},V_{44})$ est diagonale et $V_{11}$ respectivement $V_{33} > 0$, ces deux hypothèses sont vérifiées.}
\end{rem}

Pour tout $\varepsilon \in \mathbb{R}^{\ast} \setminus \mathcal{E}$, le Théorème \ref{theo3,32} montre en particulier qu'il y a une accumulation des résonances de l'opérateur $D_{\varepsilon} := D_{0} + \varepsilon V$ près des seuils $\pm m$. Pour $\pm V < 0$, les seules résonances sont les valeurs propres $z_{\pm m}(k)$ avec $k \in \textup{e}^{i\frac{\pi}{2}} ]0,\eta[$ et $\vert k \vert$ suffisamment petit. Pour $\pm V > 0$, il n'y a pas d'accumulation de valeurs propres près de $\pm m$. Nous pouvons comparer nos résultats à ceux de \cite{tda} sur la distribution asymptotique près de $\pm m$ du spectre discret dans $(-m,m)$ de $D_{\pm} := D_{0} \pm V$. La perturbation $V$ étant telle que $V \geq 0$ et ses coefficients $V_{jk}(x)$, $1 \leq j,k \leq 4$ de l'ordre de $\mathcal{O} \left( \langle x \rangle^{-m} \right)$ avec $m > 3$. En effet dans le Théorème 6.5 de \cite{tda}, par des majorations et des minorations de la fonction de décalage spectrale associée à la paire d'opérateurs $(D_{\pm},D_{0})$, R. Tiedra de Aldecoa montre d'une part qu'il n'y a pas d'accumulation des valeurs propres dans $(-m,m)$ de $D_{\pm}$ près de $\pm m$. D'autre part, qu'il y a une accumulation des valeurs propres dans $(-m,m)$ de $D_{\mp}$ près de $\pm m$.

\section{Définitions et résultats auxiliaires}

\hspace{-0.54cm} 3.1. Dans cette sous-section, nous résumons quelques résultats dus à G. D. Raikov sur les opérateurs de type Toeplitz (voir \cite{raik}). 

\begin{lem}\label{lem1,1} 
\textup{(Lemme 2.4 de \cite{raik})} Soit $U \in L^{q}(\mathbb{R}^{2})$, $q \in [1,\infty)$. Supposons que $b$ soit un champ électromagnétique admissible (voir Introduction). Soit $p$ la projection orthogonale sur $\textup{ker} \hspace{0.5mm} H_{12}^{-}$ défini par \eqref{eq1.161}. Alors $p U p \in S_{q}(L^{2}(\mathbb{R}^{2}))$ avec
\begin{equation}\label{eq1,18} 
\Vert p U p \Vert_{q}^{q} \leq \frac{b_{0}}{2\pi} \textup{e}^{2 \hspace{0.2mm} \textup{osc} \hspace{0.5mm} \tilde{\varphi}} \Vert U \Vert_{L^{q}}^{q},
\end{equation}
où $\textup{osc} \hspace{0.5mm} \tilde{\varphi}$ est défini par \eqref{eq00}.
\end{lem}

Soit $U$ une fonction de $L^{\infty}(\mathbb{R}^{2})$. La distribution asymptotique des valeurs propres des opérateurs de type Toeplitz $p U p$ fait l'objet des trois lemmes ci-dessous. Dans le premier lemme, une densité d'états intégrés (DEI) pour l'opérateur $H_{12}^{-}$ dans $L^{2}(\mathbb{R}^{3})$ est définie comme suit: soit $\chi_{T,x_{12}}$ la fonction caractéristique du carré $x_{12} + \left( -\frac{T}{2},\frac{T}{2} \right)^{2}$ avec $x_{12} \in \mathbb{R}^{2}$ et $T > 0$. Désignons par $\mathbb{P}_{I}(H_{12}^{-})$ la projection spectrale de l'opérateur $H_{12}^{-}$ associée au borélien $I \subset \mathbb{R}$. Une fonction décroissante $\varrho : \mathbb{R} \longrightarrow [0,\infty)$ est appelée une DEI pour l'opérateur $H_{12}^{-}$ si elle satisfait pour tout point $x_{12} \in \mathbb{R}^{2}$ $$\varrho(t) = \lim_{T\rightarrow\infty} T^{-2} \hspace{0.5mm} \textup{Tr} \hspace{0.5mm} [\chi_{T,x_{12}} \mathbb{P}_{(-\infty,t)}(H_{12}^{-}) \chi_{T,x_{12}}]$$ en chaque point de continuité $t$ de $\varrho$ (voir p. e. \cite{raik}). Si $b = b_{0}$, $i.e.$ $\tilde{b} = 0$ (champ magnétique constant), nous avons $\varrho(t) = \frac{b_{0}}{2\pi} \sum_{q=0}^{\infty} \chi_{\mathbb{R}_{+}} (t - 2 b_{0} q)$, $t \in \mathbb{R}$ et $\chi_{\mathbb{R}_{+}}$ est la fonction indicatrice de $\mathbb{R}_{+}$.

\begin{lem}\label{lem2,13} 
\textup{(Lemme 3.3 de \cite{raik})} Soit $U \in C^{1}(\mathbb{R}^{2})$ telle que $$0 \leq U(x_{12}) \leq C_{1} \langle x_{12} \rangle^{-\alpha}, \hspace{0.2cm} \vert \nabla U(x_{12}) \vert \leq C_{1} \langle x_{12} \rangle^{-\alpha-1}, \hspace{0.2cm} x_{12} \in \mathbb{R}^{2},$$ avec $\alpha > 0$ et $C_{1} > 0$. Supposons que $U(x_{12}) = u_{0}(x_{12} / \vert x_{12} \vert) \vert x_{12} \vert^{-\alpha} \big{(} 1 + o(1) \big{)}$ quand $\vert x_{12} \vert \rightarrow \infty$, où $u_{0}$ est une fonction continue sur la sphère $\mathbb{S}^{1}$ non identiquement nulle, et qu'il existe une DEI pour l'opérateur $H_{12}^{-}$. Alors $$n_{+} (s,p U p) = C_{\alpha} s^{-2/\alpha} \big{(} 1 + o(1) \big{)}, \hspace{0.2cm} s \searrow 0,$$ avec 
\begin{equation}\label{eq2,24}
C_{\alpha} := \frac{b}{4\pi} \int_{\mathbb{S}^{1}} u_{0}(t)^{2/\alpha} dt.
\end{equation}
\end{lem}

\begin{lem}\label{lem2,14} 
\textup{(Lemme 3.4 de \cite{raik})} Soit $0 \leq U \in L^{\infty}(\mathbb{R}^{2})$. Supposons que $$\ln U(x_{12}) = -\mu \vert x_{12} \vert^{2\beta} \big{(} 1 + o(1) \big{)}, \hspace{0.2cm} \vert x_{12} \vert \rightarrow \infty,$$ avec $\beta > 0$ et $\mu > 0$. Alors $$n_{+} (s,p U p) = \varphi_{\beta}(s) \big{(} 1 + o(1) \big{)}, \hspace{0.2cm} s \searrow 0,$$ où l'on a posé pour $0 < s < \textup{e}^{-1}$
$$\varphi_{\beta}(s) :=
 \begin{cases}
 \frac{1}{2} b_{0} \mu^{-1/\beta} \vert \ln s \vert^{1/\beta} & \text{si } 0 < \beta < 1,\\
 \dfrac{1}{\ln(1 + 2\mu/b_{0})} \vert \ln s \vert & \text{si } \beta = 1,\\
 \dfrac{\beta}{\beta - 1} \big{(} \ln \vert \ln s \vert \big{)}^{-1} \vert \ln s \vert & \text{si } \beta > 1.
 \end{cases}$$
\end{lem}

\begin{lem}\label{lem2,15} 
\textup{(Lemme 3.5 de \cite{raik})} Soit $0 \leq U \in L^{\infty}(\mathbb{R}^{2})$. Supposons que $U$ soit à support compact et qu'il existe une constante $C > 0$ telle que $U \geq C$ sur un sous-ensemble ouvert non vide de $\mathbb{R}^{2}$. Alors $$n_{+} (s,p U p) = \varphi_{\infty}(s) \big{(} 1 + o(1) \big{)}, \hspace{0.2cm} s \searrow 0,$$ avec $$\varphi_{\infty}(s) := \big{(} \ln \vert \ln s \vert \big{)}^{-1} \vert \ln s \vert, \hspace{0.2cm} 0 < s < \textup{e}^{-1}.$$
\end{lem}

\hspace{-0.54cm} 3.2. Dans cette sous-section, nous résumons des résultats dus à J. F. Bony, V. Bruneau et G. D. Raikov sur les valeurs caractéristiques d'une fonction holomorphe à valeur opérateur (voir \cite{bon} et \cite{bo}).

\hspace{-0.54cm} Le lemme suivant est une conséquence de l'inégalité classique de Jensen.

\begin{lem}\label{lem2,17} 
\textup{(Lemme 6 de \cite{bon})} Soient $\Delta$ un domaine simplement connexe de $\mathbb{C}$, et $g$ une fonction holomorphe sur $\Delta$ ayant une extension sur $\overline{\Delta}$ continue. Supposons qu'il existe $\lambda_{0} \in \Delta$ tel que $g(\lambda_{0}) \neq 0$, et $g(\lambda) \neq 0$ pour $\lambda$ appartenant à la frontière $\partial \Delta$ de $\Delta$. Soient $\lambda_{1}, \lambda_{2}, \ldots, \lambda_{N} \in \Delta$ les zéros de $g$ répétés avec leur multiplicité. Alors pour tout domaine $\Delta' \Subset \Delta$, il existe une constante $C' > 0$ telle que $N(\Delta',g)$, le nombre de zéros $\lambda_{j}$ de $g$ contenus dans $\Delta'$, vérifie
\begin{equation}\label{eq2,32}
N(\Delta',g) \leq C' \left( \int_{\partial \Delta} \textup{ln} \vert g(\lambda) \vert d\lambda - \textup{ln} \vert g(\lambda_{0}) \vert  \right).
\end{equation}
\end{lem}

Soit un domaine $\mathcal{D}$ de $\mathbb{C}$ contenant $0$ et considérons une fonction holomorphe à valeur opérateur compact $T : \mathcal{D} \longrightarrow S_{\infty}$. Pour un domaine $\Omega \subset \mathcal{D} \setminus \lbrace 0 \rbrace$, un nombre complexe $z \in \Omega$ est une \textit{valeur caractéristique} de $z \mapsto I - \frac{T(z)}{z}$ si l'opérateur $I - \frac{T(z)}{z}$ n'est pas inversible. Désignons par $$\mathcal{Z}(\Omega) := \left\lbrace z \in \Omega : I - \frac{T(z)}{z} \hspace{0.8mm} \textup{n'est pas inversible} \right\rbrace.$$ Dès qu'il existe $z_{0} \in \Omega$ tel que l'opérateur $I - \frac{T(z_{0})}{z_{0}}$ est inversible, alors l'ensemble $\mathcal{Z}(\Omega)$ est discret (voir p. e. la Proposition 4.1.4 de \cite{go}). On définit alors $$\mathcal{N}(\Omega) := \# \mathcal{Z}(\Omega).$$ 
Supposons que $T(0)$ soit auto-adjoint. Soient $\Omega \Subset \mathbb{C} \setminus \lbrace 0 \rbrace$ et 
\begin{equation}\label{eq2,33}
\mathcal{C}_{\alpha}(a,b) := \lbrace x + iy \in \mathbb{C} : a \leq x \leq b, -\alpha x \leq y \leq \alpha x \rbrace
\end{equation}
avec $a > 0$ tendant vers $0$ et $b > 0$. Soit $$n(\Lambda) := \textup{Tr} \hspace{0.6mm} \mathbf{1}_{\Lambda}(T(0))$$ le nombre de valeurs propres comptées avec leur multiplicité de l'opérateur $T(0)$ dans $\Lambda$ un intervalle de $\mathbb{R}^{\ast}$. Désignons par $\Pi_{0}$ la projection orthogonale sur $\textup{ker} \hspace{0.6mm} T(0)$.

\begin{lem}\label{lem2,19} 
\textup{(Corollaire 3.4 de \cite{bo})} Soit $T$ comme ci-dessus et tel que $I - T'(0) \Pi_{0}$ soit inversible. Supposons que $\Omega \Subset \mathbb{C} \setminus \lbrace 0 \rbrace$ soit un domaine borné de frontière régulière qui soit transverse à l'axe réel en chaque point de leur intersection.

$\textup{(i)}$ Si $\Omega \cap \mathbb{R} = \emptyset$, alors $\mathcal{N}(s\Omega) = 0$ pour $s$ assez petit. Ceci entraîne que les valeurs caractéristiques $z \in \mathcal{Z}(\mathcal{D})$ près de $0$ vérifient $\vert \textup{Im} \hspace{0.6mm} z \vert = o(\vert z \vert)$.

$\textup{(ii)}$ De plus, si l'opérateur $T(0)$ est de signe défini $(\pm T(0) \geq 0)$, près de $0$ les valeurs caractéristiques vérifient $\pm \textup{Re} \hspace{0.6mm} z \geq 0$, respectivement.

$\textup{(iii)}$ Si $T(0)$ est de rang fini, il n'y a pas de valeur caractéristique dans un voisinage de $0$. De plus, si l'opérateur $T(0) \mathbf{1}_{[0,+\infty)}(\pm T(0))$ est de rang fini, il n'y a pas de valeur caractéristique dans un voisinage de $0$ intersectant $\lbrace \pm \textup{Re} \hspace{0.6mm} z > 0 \rbrace$, respectivement.
\end{lem}

\begin{lem}\label{lem2,20} 
\textup{(Corollaire 3.9 de \cite{bo})} Soit $T$ comme ci-dessus et tel que $I - T'(0) \Pi_{0}$ soit inversible. Supposons qu'il existe $\gamma > 0$ tel que $$n([r,1]) = O(r^{-\gamma}), \hspace{0.2cm} r \searrow 0,$$ et que $n([r,1])$ croît vers l'infini quand $r \searrow 0$. Alors il existe une suite positive $(r_{k})_{k}$ qui tend vers $0$ telle que 
\begin{equation}\label{eq2,34}
\mathcal{N}(\mathcal{C}_{\alpha}(r_{k},1)) = n([r_{k},1]) (1 + o(1)), \hspace{0.2cm} k \rightarrow \infty.
\end{equation}
\end{lem}

\begin{lem}\label{lem2,21} 
\textup{(Corollaire 3.11 de \cite{bo})} Soit $T$ comme ci-dessus et tel que $I - T'(0) \Pi_{0}$ soit inversible. Supposons que $$n([r,1]) = \Phi(r)(1 + o(1)), \hspace{0.2cm} r \searrow 0,$$ avec $\Phi(r) = \vert \ln r \vert^{\gamma}$, $\gamma > 0$, ou $\Phi(r) = \big{(} \ln \vert \ln r \vert \big{)}^{-1} \vert \ln r \vert$, ou encore $\Phi(r) = r^{-\gamma}$, $\gamma > 0$. Alors
\begin{equation}\label{eq2,35}
\mathcal{N}(\mathcal{C}_{\alpha}(r,1)) = \Phi(r) (1 + o(1)), \hspace{0.2cm} r \searrow 0.
\end{equation}
\end{lem}

\section{Résonances près de $0$ de l'opérateur de Pauli}

Nous supposons que la perturbation $V$ vérifie (pour $n = 2$) l'hypothèse \eqref{eq1,14} dans toute la section.

\subsection{Résonances}

Soit $H_{0}$ l'opérateur de Pauli défini par \eqref{eq1,9}. Dans cette partie, nous définissons les résonances près de $0$ de l'opérateur perturbé $H := H_{0} + V$.

Soient $P := p \otimes 1$ et $Q := I - P$, où $p$ est la projection orthogonale sur $\textup{ker} \hspace{0.5mm} H_{12}^{-}$ défini par \eqref{eq1.161}. Considérons également les projections orthogonales dans $L^{2}(\mathbb{R}^{3},\mathbb{C}^{2})$
\begin{equation}\label{eq2,4}
\textup{P} := \begin{pmatrix} 
   P & 0 \\ 
   0 & 0 
\end{pmatrix}, \hspace{1cm} \textup{Q} := \textup{I} - \textup{P} = \begin{pmatrix} 
   Q & 0 \\
   0 & I 
\end{pmatrix}.
\end{equation}
Soit $z \in \mathbb{C} \setminus \sigma (H_{0})$. Via (\ref{eq1,10}), nous avons
\begin{equation}\label{eq2,8}
\left( H_{0} - z \right)^{-1} \textup{P} = \begin{pmatrix}
p \otimes \mathcal{R}(z) & 0 \\
   0 & 0
\end{pmatrix},
\end{equation}
où la résolvante $\mathcal{R}(z) := \left( \Pi_{3}^{2} - z \right)^{-1}$ bornée dans $L^{2}(\mathbb{R})$ admet pour noyau intégral 
\begin{equation}\label{eq2,81}
\mathcal{N}_{z}(x_{3} - x'_{3}) = i \textup{e}^{i \sqrt{z} \vert x_{3} - x'_{3} \vert} / (2 \sqrt{z}).
\end{equation}
Par conséquent
\begin{equation}\label{eq2,9}
\left( H_{0} - z \right)^{-1} = \big{[} p \otimes \mathcal{R}(z) \big{]} \begin{pmatrix}
1 & 0\\
0 & 0
\end{pmatrix} + \left( H_{0} - z \right)^{-1} \textup{Q}.
\end{equation}
Soit $z \in \mathbb{C}^{+}$. Effectuons le changement de variables
\begin{equation}\label{eq2,10}
z := z(k) = k^{2} \hspace{0.2cm} \textup{pour} \hspace{0.2cm} k \in \mathbb{C}_{1/2}^{+} := \lbrace k \in \mathbb{C}^{+} : k^{2} \in \mathbb{C}^{+} \rbrace.
\end{equation}

La première étape consiste à prolonger holomorphiquement en la variable $k \in \mathbb{C}_{1/2}^{+}$ l'opérateur $\left( H_{0} - z(k) \right)^{-1} \textup{P}$ au voisinage de $k = 0$. Puisque le spectre de $H_{0}$ est $[0,+\infty)$, le prolongement pourrait se faire à partir de $k \in \mathbb{C}^{+}$ \big{(}$\Leftrightarrow k^{2} \in \mathbb{C} \setminus [0,+\infty)$\big{)}. Cependant, $H$ peut avoir des valeurs propres négatives. Nous prolongeons donc $\left( H - z(k) \right)^{-1}$ à partir de $k \in \mathbb{C}_{1/2}^{+}$ tout en sachant que ses seuls pôles dans $\mathbb{C}^{+}$ sont sur la demi-droite $i]0,+\infty)$.
\begin{lem}\label{prop2,5} 
Soient $\delta$ et $D(0,\epsilon)^{\ast}$ définis respectivement par \eqref{eq1,14} et \eqref{eq2,110}.

$\textup{(i)}$ L'application $$k \longmapsto \left( \left( H_{0} - z(k) \right)^{-1} \textup{P} : \textup{e}^{-\delta \langle x_{3} \rangle} L^{2}(\mathbb{R}^{3}) \longrightarrow \textup{e}^{\delta \langle x_{3} \rangle} L^{2}(\mathbb{R}^{3}) \right)$$ admet un prolongement holomorphe de $\mathbb{C}_{1/2}^{+} \cap D(0,\epsilon)^{\ast}$ à $D(0,\epsilon)^{\ast}$ avec $\epsilon < \delta$. 

$\textup{(ii)}$ Soit $v_{12}(x_{12}) := \langle x_{12} \rangle^{-\alpha}$, $\alpha > 0$. L'application $$T_{v_{12}} : k \longmapsto v_{12}(x_{12}) \textup{e}^{-\delta \langle x_{3} \rangle} \left( H_{0} - z(k) \right)^{-1} \textup{P} \textup{e}^{-\delta \langle x_{3} \rangle}$$ se prolonge aussi holomorphiquement sur $D(0,\epsilon)^{\ast}$, à valeurs dans $S_{\infty} \left( L^{2}(\mathbb{R}^{3}) \right)$.
\end{lem}

\hspace{-0.54cm} \begin{preuve}
(i) Vu comme opérateur de $\textup{e}^{-\delta \langle x_{3} \rangle} L^{2}(\mathbb{R}^{3})$ dans $\textup{e}^{\delta \langle x_{3} \rangle} L^{2}(\mathbb{R}^{3})$, soit 
$$L(k) = \big{[} p \otimes \mathcal{R}(k^{2}) \big{]} \begin{pmatrix}
1 & 0\\
0 & 0
\end{pmatrix}.$$
Le noyau intégral de l'opérateur $N(k) := \textup{e}^{-\delta \langle x_{3} \rangle} \mathcal{R}(k^{2}) \textup{e}^{-\delta \langle x_{3} \rangle}$ est  
\begin{equation}\label{eq2,91}
\textup{e}^{-\delta \langle x_{3} \rangle} \frac{i \textup{e}^{i k \vert x_{3} -x'_{3} \vert}}{2 k} \textup{e}^{-\delta \langle x'_{3} \rangle}.
\end{equation}
Il est dans $L^{2}(\mathbb{R}^{2})$ pour $\textup{Im} \hspace{0.5mm} k > - \delta$. Donc si $\epsilon < \delta$, on peut prolonger holomorphiquement $k \longmapsto L(k) \in \mathcal{L} \left( \textup{e}^{-\delta \langle x_{3} \rangle} L^{2}(\mathbb{R}^{3}),\textup{e}^{\delta \langle x_{3} \rangle} L^{2}(\mathbb{R}^{3}) \right)$ de $\mathbb{C}_{1/2}^{+} \cap D(0,\epsilon)^{\ast}$ à $D(0,\epsilon)^{\ast}$. D'où $k \longmapsto \left( H_{0} - z(k) \right)^{-1} \textup{P} \in \mathcal{L} \left( \textup{e}^{-\delta \langle x_{3} \rangle} L^{2}(\mathbb{R}^{3}),\textup{e}^{\delta \langle x_{3} \rangle} L^{2}(\mathbb{R}^{3}) \right)$ admet un prolongement holomorphe à $D(0,\epsilon)^{\ast}$.

(ii) Par (\ref{eq2,8})
\begin{equation}\label{eq2,12}
T_{v_{12}}(k) = \left[ v_{12}p \otimes N(k) \right] \begin{pmatrix}
1 & 0\\
0 & 0
\end{pmatrix}.
\end{equation}
De la preuve de (i), il découle que $N(k) \in S_{2}$ dans $L^{2}(\mathbb{R})$ pour $\textup{Im} \hspace{0.5mm} k > - \delta$. Le Lemme \ref{lem1,1} entraîne que $p v_{12}^{2} p$ est dans une certaine classe $S_{q}$ pour $q$ tel que $\alpha q > 1$; d'où sa compacité. Ce qui est équivalent à la compacité de l'opérateur $v_{12} p$. Donc $T_{v_{12}}(k)$ est compact pour $\textup{Im} \hspace{0.5mm} k > - \delta$ et $k \mapsto T_{v_{12}}(k) \in S_{\infty} \left( L^{2}(\mathbb{R}^{3}) \right)$ se prolonge comme ci-dessus holomorphiquement de $\mathbb{C}_{1/2}^{+} \cap D(0,\epsilon)^{\ast}$ à $D(0,\epsilon)^{\ast}$. D'où le lemme.
\end{preuve}

La deuxième étape consiste à prolonger holomorphiquement l'opérateur $\left( H_{0} - z \right)^{-1} \textup{Q}$ en la variable $z$.

\begin{lem}\label{prop2,6} 
Soient $\delta$ et $\zeta$ définis respectivement par \eqref{eq1,14} et \eqref{eq0}.

$\textup{(i)}$ l'application $$z \longmapsto \left( \left( H_{0} - z \right)^{-1} \textup{Q} : \textup{e}^{-\delta \langle x_{3} \rangle} L^{2}(\mathbb{R}^{3}) \longrightarrow \textup{e}^{\delta \langle x_{3} \rangle} L^{2}(\mathbb{R}^{3}) \right)$$ admet un prolongement holomorphe de $\mathbb{C}^{+}$ à $\mathbb{C} \setminus [\zeta,\infty)$.

$\textup{(ii)}$ Soit $v_{12}(x_{12}) := \langle x_{12} \rangle^{-\alpha}$, $\alpha > 0$. L'application $$L_{v_{12}} : z \longmapsto v_{12}(x_{12}) \textup{e}^{-\delta \langle x_{3} \rangle} \left( H_{0} - z \right)^{-1} \textup{Q} \textup{e}^{-\delta \langle x_{3} \rangle}$$ se prolonge aussi holomorphiquement à $\mathbb{C} \setminus [\zeta,\infty)$, à valeurs dans $S_{\infty} \left( L^{2}(\mathbb{R}^{3}) \right)$.
\end{lem}

\hspace{-0.54cm} \begin{preuve}
(i) Soit $z \in \mathbb{C}^{+}$. Par (\ref{eq1,10}) l'opérateur $\left( H_{0} - z \right)^{-1} \textup{Q}$ est donné par
\begin{equation}\label{eq2,13}
\left( \begin{smallmatrix}
   (H_{0}^{-} - z)^{-1} Q & 0\\
   0 & (H_{0}^{+} - z)^{-1}
\end{smallmatrix} \right) = (H_{0}^{-} - z)^{-1} Q \oplus (H_{0}^{+} - z)^{-1}.
\end{equation}
Ainsi, $\mathbb{C} \setminus [\zeta,\infty) \ni z \longmapsto (H_{0}^{-} - z)^{-1} Q \oplus (H_{0}^{+} - z)^{-1}$ est bien définie et analytique car $\mathbb{C} \setminus [\zeta,\infty)$ est contenu dans l'ensemble résolvent de l'opérateur $H_{0}^{-}$ restreint à $Q D(H_{0}^{-})$, et de $H_{0}^{+}$ restreint à $D(H_{0}^{+})$. L'opérateur $\textup{e}^{-\delta \langle x_{3} \rangle} \left( H_{0} - z \right)^{-1} \textup{Q} \textup{e}^{-\delta \langle x_{3} \rangle}$ se prolonge donc holomorphiquement à $\mathbb{C} \setminus [\zeta,\infty)$.

(ii) L'identité (\ref{eq2,13}) entraîne que $$L_{v_{12}}(z) = v_{12} \textup{e}^{-\delta \langle x_{3} \rangle} (H_{0}^{-} - z)^{-1} Q \textup{e}^{-\delta \langle x_{3} \rangle} \oplus v_{12} \textup{e}^{-\delta \langle x_{3} \rangle} (H_{0}^{+} - z)^{-1} \textup{e}^{-\delta \langle x_{3} \rangle}.$$ En utilisant l'inégalité diamagnétique (Théorème 2.3 de \cite{avr}) et le Théorème 2.13 de \cite{sim}, on montre en raisonnant comme dans la preuve de la Proposition 4.4 de \cite{raik} que l'opérateur $L_{v_{12}}(z)$ est dans une certaine classe $S_{q}$, pour $q$ paire tel que $q > 3$ et $\alpha q > 2$. Donc $L_{v_{12}}(z)$ est dans $S_{\infty} \left( L^{2}(\mathbb{R}^{3}) \right)$ et se prolonge holomorphiquement de $\mathbb{C}^{+}$ à $\mathbb{C} \setminus [\zeta,\infty)$.
\end{preuve}

Le lemme suivant découle directement des Lemmes \ref{prop2,5} et \ref{prop2,6}.

\begin{lem}\label{lem2,9} 
Soit $D(0,\epsilon)^{\ast}$ défini par \eqref{eq2,110}. Supposons que $V$ vérifie \eqref{eq1,14} pour $n = 2$. L'application $$\mathbb{C}_{1/2}^{+} \cap D(0,\epsilon)^{\ast} \ni k \longmapsto \mathcal{T}_{V}(z(k)) := J \vert V \vert^{1/2} \left( H_{0} - z(k) \right)^{-1} \vert V \vert^{1/2},$$ avec $J := \textup{sign} \hspace{0.5mm} V$, se prolonge analytiquement à $D(0,\epsilon)^{\ast}$ à valeurs dans $S_{\infty} \left( L^{2}(\mathbb{R}^{3}) \right)$. Ce prolongement est encore noté $\mathcal{T}_{V}(z(k))$.
\end{lem}

\hspace{-0.54cm} L'identité $$\left( H - z \right)^{-1} \left( 1 + V(H_{0} - z)^{-1} \right) = (H_{0} - z)^{-1}$$ donne
\begin{align*}
\textup{e}^{-\delta \langle x_{3} \rangle} \left( H - z \right)^{-1} \textup{e}^{-\delta \langle x_{3} \rangle} & = \textup{e}^{-\delta \langle x_{3} \rangle} (H_{0} - z)^{-1} \textup{e}^{-\delta \langle x_{3} \rangle} \\ 
& \times \left( 1 + \textup{e}^{\delta \langle x_{3} \rangle} V(H_{0} - z)^{-1} \textup{e}^{-\delta \langle x_{3} \rangle} \right)^{-1}.
\end{align*}
Par le Lemme \ref{lem2,9}, la fonction $k \longmapsto \textup{e}^{\delta \langle x_{3} \rangle}V(H_{0} - z(k))^{-1} \textup{e}^{-\delta \langle x_{3} \rangle}$ est holomorphe à valeur opérateur compact dans $L^{2}(\mathbb{R}^{3})$ inversible en au moins un point. D'où par le théorème analytique de Fredholm, $$k \longmapsto \left( 1 + \textup{e}^{\delta \langle x_{3} \rangle} V(H_{0} - z(k))^{-1} \textup{e}^{-\delta \langle x_{3} \rangle} \right)^{-1}$$ admet un prolongement méromorphe de $\mathbb{C}_{1/2}^{+} \cap D(0,\epsilon)^{\ast}$ à $D(0,\epsilon)^{\ast}$. Ce qui nous permet de définir les résonances de $H$ près de $0$.

\begin{prop}\label{prop2,8} 
Soient $\mathbb{C}_{1/2}^{+}$ défini par \eqref{eq2,10} et $D(0,\epsilon)^{\ast}$ par \eqref{eq2,110}. L'application $$k \longmapsto \left( \left( H - z(k) \right)^{-1} : \textup{e}^{-\delta \langle x_{3} \rangle} L^{2}(\mathbb{R}^{3}) \longrightarrow \textup{e}^{\delta \langle x_{3} \rangle} L^{2}(\mathbb{R}^{3}) \right)$$ admet un prolongement méromorphe de $\mathbb{C}_{1/2}^{+} \cap D(0,\epsilon)^{\ast}$ à $D(0,\epsilon)^{\ast}$. Ce prolongement est encore noté $R(z(k))$.
\end{prop}

La notion d'indice (le long d'un contour fermé orienté positivement) d'une fonction méromorphe finie à valeur opérateur qui est définie dans l'appendice, permet de définir la multiplicité d'une résonance.

\begin{déf}\label{def2,11} 
\textup{Nous définissons les résonances de l'opérateur $H$ (près de $0$) comme étant les pôles du prolongement méromorphe noté $R(z)$, de la résolvante $\left( H - z \right)^{-1}$ dans $\mathcal{L} \left( \textup{e}^{-\delta \langle x_{3} \rangle} L^{2}(\mathbb{R}^{3}),\textup{e}^{\delta \langle x_{3} \rangle} L^{2}(\mathbb{R}^{3}) \right)$. La multiplicité d'une résonance $z_{1} := z(k_{1}) = k_{1}^{2}$ est définie par 
\begin{equation}\label{eq2,14}
\textup{mult}(z_{1}) := \textup{Ind}_{\gamma} \hspace{0.5mm} \left( I + \mathcal{T}_{V}(z(\cdot)) \right),
\end{equation}
où $\gamma$ est un cercle assez petit orienté positivement contenant $k_{1}$ comme unique point vérifiant que $z(k_{1})$ est résonance de $H$, et $\mathcal{T}_{V}(z(\cdot))$ est défini au Lemme \ref{lem2,9}.}
\end{déf}

\begin{rem}\label{rem2,70} 
\textup{Si $H_{0}$ est l'opérateur de Schrödinger magnétique défini dans \cite{bon}, l'opérateur $\mathcal{T}_{V}(z)$ est dans $S_{2}$ et $\partial_{z} \mathcal{T}_{V}(z)$ est dans $S_{1}$. Et dans ce cas la Définition \ref{def2,11} coïncide avec la Définition 3 de \cite{bon} (voir aussi la Définition 4.3 de \cite{sj}).}
\end{rem} 

\begin{prop}\label{prop2,10} 
Pour $k$ proche de $0$, les assertions suivantes sont équivalentes:

$\textup{(i)}$ $z(k) = k^{2}$ est un pôle de $R(z(k))$,

$\textup{(ii)}$ $-1$ est une valeur propre de $\mathcal{T}_{V}(z(k)) := J \vert V \vert^{1/2} R_{0}(z(k)) \vert V \vert^{1/2}$.
\end{prop}

\hspace{-0.54cm} \begin{preuve}
Ce résultat découle directement de l'identité
\begin{equation}\label{eq2,131}
\left( I + J \vert V \vert^{1/2} R_{0}(z) \vert V \vert^{1/2} \right) \left( I - J \vert V \vert^{1/2} R(z) \vert V \vert^{1/2} \right) = I.
\end{equation}
\end{preuve}

\subsection{Preuve du Théorème \ref{theo2,18}}

Rappelons que $p$ est la projection orthogonale sur $\textup{ker} \hspace{0.5mm} H_{12}^{-}$ défini par \eqref{eq1.161}. Décomposons grâce à \eqref{eq2,9} l'opérateur $\mathcal{T}_{V}(z(k))$ (défini au Lemme \ref{lem2,9}) de la manière suivante: $\mathcal{T}_{V}(z(k)) = \mathcal{T}_{1}^{V}(k) + \mathcal{T}_{2}^{V}(k)$ où
$$\mathcal{T}_{1}^{V}(k) := J \vert V \vert^{1/2} \left[ p \otimes \mathcal{R} ( k^{2} ) \right] \begin{pmatrix}
   1 & 0 \\
   0 & 0
\end{pmatrix} \vert V \vert^{1/2},$$
$$\mathcal{T}_{2}^{V}(k) := J \vert V \vert^{1/2} \left( H_{0} - z(k) \right)^{-1} \textup{Q} \vert V \vert^{1/2}.$$
Par le Lemme \ref{prop2,6}, l'opérateur $\mathcal{T}_{2}^{V}(k)$ est holomorphe dans un voisinage de $0$ à valeurs dans $S_{\infty} \left( L^{2}(\mathbb{R}^{3}) \right)$. Considérons à présent $\mathcal{T}_{1}^{V}(k)$ pour $k \in D(0,\epsilon)^{\ast}$ défini par \eqref{eq2,110}. Le noyau intégral de l'opérateur $N(k) := \textup{e}^{-\delta \langle x_{3} \rangle} \mathcal{R} (k^{2}) \textup{e}^{-\delta \langle x_{3} \rangle}$ est donné par \eqref{eq2,91}. Ce qui nous permet d'écrire
\begin{equation}\label{eq2,15}
N(k) = \frac{1}{k}t_{1} + r_{1}(k),
\end{equation}
où $t_{1} : L^{2}(\mathbb{R}) \longrightarrow L^{2}(\mathbb{R})$ est l'opérateur de rang 1 défini par 
\begin{equation}\label{eq2,16}
t_{1}(u) := \frac{i}{2} \langle u,\textup{e}^{-\delta \langle \cdot \rangle} \rangle \textup{e}^{-\delta \langle x_{3} \rangle},
\end{equation}
et $r_{1}(k)$ est l'opérateur de Hilbert-Schmidt (sur $D(0,\epsilon)^{\ast}$) ayant pour noyau intégral 
\begin{equation}\label{eq2,17}
\textup{e}^{-\delta \langle x_{3} \rangle} i \frac{ \textup{e}^{ i k \vert x_{3} -x'_{3} \vert} - 1}{2 k} \textup{e}^{-\delta \langle x'_{3} \rangle}.
\end{equation}
Donc 
\begin{equation}\label{eq2,18}
\begin{aligned}
\mathcal{T}_{1}^{V}(k) = \frac{iJ}{k} \times \frac{1}{2} \vert V \vert^{1/2} & \left[ p \otimes \tau_{1} \right] \begin{pmatrix}
   1 & 0 \\
   0 & 0
\end{pmatrix} \vert V \vert^{1/2} \\
& + J \vert V \vert^{1/2} \left[ p \otimes s_{1}(k) \right] \begin{pmatrix}
   1 & 0 \\
   0 & 0
\end{pmatrix} \vert V \vert^{1/2},
\end{aligned}
\end{equation}
où $\tau_{1}$ et $s_{1}(k)$ sont des opérateurs dans $L^{2}(\mathbb{R})$ ayant respectivement pour noyau intégral $1$ et
\begin{equation}\label{eq2,19}
\frac{ 1 - \textup{e}^{ i k \vert x_{3} -x'_{3} \vert}}{2 i k}.
\end{equation}
Soit l'opérateur borné $K : L^{2}(\mathbb{R}^{3},\mathbb{C}^{2}) \rightarrow L^{2}(\mathbb{R}^{2},\mathbb{C}^{2})$ défini par 
\begin{equation}\label{eq2,190}
(K \psi)(x_{12}) := \int_{\mathbb{R}^{3}} \mathcal{P}(x_{12},x'_{12}) \begin{pmatrix}
   1 & 0 \\
   0 & 0
\end{pmatrix} \vert V \vert^{1/2}(x'_{12},x'_{3})\psi(x'_{12},x'_{3}) dx'_{12}dx'_{3},
\end{equation}
où $\mathcal{P}(\cdot,\cdot)$ est le noyau intégral de la projection $p$. Son adjoint\\ $K^{\ast} : L^{2}(\mathbb{R}^{2},\mathbb{C}^{2}) \rightarrow L^{2}(\mathbb{R}^{3},\mathbb{C}^{2})$ est donné par
$$(K^{\ast} \varphi)(x_{12},x_{3}) = \vert V \vert^{1/2}(x_{12},x_{3}) \begin{pmatrix}
   1 & 0 \\
   0 & 0
\end{pmatrix} (p\varphi)(x_{12})$$
et nous avons
$$K^{\ast} K = \vert V \vert^{1/2} [p \otimes \tau_{1}] \begin{pmatrix}
   1 & 0 \\
   0 & 0
\end{pmatrix} \vert V \vert^{1/2}.$$
Finalement (\ref{eq2,18}) devient
$$\mathcal{T}_{1}^{V}(k) = \frac{iJ}{k} \times \frac{1}{2} K^{\ast} K + J \vert V \vert^{1/2} \left[ p \otimes s_{1}(k) \right] \begin{pmatrix}
   1 & 0 \\
   0 & 0
\end{pmatrix} \vert V \vert^{1/2}.$$
D'où la

\begin{prop}\label{prop2,12} 
Pour $k \in D(0,\epsilon)^{\ast}$,
\begin{equation}\label{eq2,20}
\mathcal{T}_{V}(z(k)) = \frac{iJ}{k} B + A(k), 
\end{equation}
où $J := \textup{sign} \hspace{0.5mm} V$, $B$ est l'opérateur auto-adjoint positif défini par
\begin{equation}\label{eq2,21}
B := \frac{1}{2} K^{\ast} K,
\end{equation} 
$K$ donné par \eqref{eq2,190}, et $A(k) \in S_{\infty} \left( L^{2}(\mathbb{R}^{3}) \right)$ est l'opérateur holomorphe sur $D(0,\epsilon)^{\ast}$ défini par
\begin{equation}\label{eq2,212}
A(k) := J \vert V \vert^{1/2} \left[ p \otimes s_{1}(k) \right] \begin{pmatrix}
   1 & 0 \\
   0 & 0
\end{pmatrix} \vert V \vert^{1/2} + \mathcal{T}_{2}^{V}(k).
\end{equation} 
\end{prop}

Soit $W$ la fonction définie par \eqref{eq2,22}. Nous avons
$$K K^{\ast} = \begin{pmatrix}
   1 & 0 \\
   0 & 0
\end{pmatrix} p W p.$$
Alors pour $s > 0$,
\begin{align*}
n_{+} \big{(} s,\frac{1}{2} K^{\ast} K \big{)} & = n_{+} \big{(} 2s,K^{\ast} K \big{)} = n_{+} \big{(} 2s,K K^{\ast} \big{)}\\
& = n_{+} \left( 2s, \left( \begin{smallmatrix}
   1 & 0 \\
   0 & 0
\end{smallmatrix} \right) p W p \right)\\
& = n_{+} \left( 2s,p W p \right).
\end{align*}
C'est-à-dire
\begin{equation}\label{eq2,23}
n_{+} \left( s,B \right) = n_{+} \left( 2s,p W p \right).
\end{equation}

\begin{prop}\label{prop2,16} 
Soit $s_{0} < \epsilon$ assez petit. Pour $0 < s < \vert k \vert < s_{0}$,
 
$\textup{(i)}$ $z(k)$ est une résonance de $H$ (près de $0$) si et seulement si $k$ est un zéro de 
\begin{equation}\label{eq2,25}
D(k,s) := \det \big{(} I + K(k,s) \big{)},
\end{equation}
où $K(k,s)$ est un opérateur de rang fini analytique en $k$, vérifiant  
$$\textup{rang} \hspace{0.6mm} K(k,s) = O \big{(} n_{+}(s,pWp) + 1 \big{)}, \hspace{0.3cm} \left\Vert K(k,s) \right\Vert = O \left( s^{-1} \right),$$
uniformément pour $s < \vert k \vert < s_{0}$.

$\textup{(ii)}$ Si $z_{1} := z(k_{1}) = k_{1}^{2}$ est une résonance de $H$, alors
\begin{equation}\label{eq2,26}
\textup{mult}(z_{1}) = \textup{Ind}_{\gamma} \hspace{0.5mm} \left( I + K(k,s) \right) = \textup{m}(k_{1}),
\end{equation}
$\gamma$ étant choisi comme dans la \textup{Définition \ref{def2,11}} et $\textup{m}(k_{1})$ désigne la multiplicité de $k_{1}$ en tant que zéro de $D(k,s)$.

$\textup{(iii)}$ Si $\textup{Im} \hspace{0.5mm} k^{2} > \varsigma > 0$, l'opérateur $I + K(k,s)$ est inversible avec 
$$\left\Vert \big{(} I + K(k,s) \big{)}^{-1} \right\Vert = O \left( \varsigma^{-1} \right),$$
uniformément pour $s < \vert k \vert < s_{0}$.
\end{prop}

\hspace{-0.54cm} \begin{preuve}
$\textup{(i)}$,$\textup{(ii)}$ Par la Proposition \ref{prop2,12}, pour $s < \vert k \vert \leq s_{0} < \epsilon$, $$\mathcal{T}_{V}(z) = \frac{iJ}{k} B + A(k).$$
L'application $k \mapsto A(k)$ est analytique près de $0$ à valeurs dans $S_{\infty}$. Donc pour $s_{0}$ assez petit, il existe un opérateur $A_{0}$ de rang fini indépendant de $k$, et $\tilde{A}(k)$ analytique près de $0$ dans $S_{\infty}$ vérifiant $\Vert \tilde{A}(k) \Vert < \frac{1}{4}$, $\vert k \vert \leq s_{0}$ tels que $$A(k) = A_{0} + \tilde{A}(k).$$ Soit la décomposition 
\begin{equation}\label{eq2,27}
B = B \mathbf{1}_{[0,\frac{1}{2}s]} (B) + B \mathbf{1}_{]\frac{1}{2}s,\infty[} (B).
\end{equation}
Puisque $\left\Vert (iJ/k) B \mathbf{1}_{[0,\frac{1}{2}s]} (B) + \tilde{A}(k) \right\Vert < \frac{3}{4}$ pour $0 < s < \vert k \vert < s_{0}$, alors 
\begin{equation}\label{eq2,28}
\big{(} I + \mathcal{T}_{V}(z) \big{)} = \big{(} I + K(k,s) \big{)} \left( I + \frac{iJ}{k} B \mathbf{1}_{[0,\frac{1}{2}s]} (B) + \tilde{A}(k) \right),
\end{equation}
où l'opérateur $K(k,s)$ est donné par
$$K(k,s) = \left( \frac{iJ}{k} B \mathbf{1}_{]\frac{1}{2}s,\infty[} (B) + A_{0} \right) \left( I + \frac{iJ}{k} B \mathbf{1}_{[0,\frac{1}{2}s]} (B) + \tilde{A}(k) \right)^{-1}.$$
Opérateur dont le rang est majoré d'une part par $$O \left( n_{+} \left( \frac{1}{2}s,B \right) + 1 \right) = O (n_{+}(s,pWp) + 1)$$
via (\ref{eq2,23}), et d'autre part dont la norme est majorée par $O \left( \vert k \vert^{-1} \right)$.\\
\\
Par ailleurs, puisque $\Vert (iJ/k) B \mathbf{1}_{[0,\frac{1}{2}s]} (B) + \tilde{A}(k)\Vert < 1$ pour $0 < s < \vert k \vert < s_{0}$, alors par le Théorème 4.4.3 de \cite{goh} $$\textup{Ind}_{\gamma} \hspace{0.5mm} \left(I + (iJ/k) B \mathbf{1}_{[0,\frac{1}{2}s]} (B) + \tilde{A}(k) \right) = 0.$$ En appliquant à (\ref{eq2,28}) les propriétés sur l'indice d'un opérateur sur un contour énoncées dans l'appendice, nous obtenons facilement les égalités (\ref{eq2,26}). Par conséquent, par la Définition \ref{def2,11} et la Proposition \ref{prop2,10} combinées à (\ref{eq2,28}), on voit que $z := z(k)$ est une résonance de $H$ si et seulement si $k$ est un zéro du déterminant $D(k,s)$ défini par (\ref{eq2,25}).

$\textup{(iii)}$ Pour $0 < s < \vert k \vert < s_{0}$,
$$I + K(k,s) = (I + \mathcal{T}_{V}(z(k))) \left( I + \frac{iJ}{k} B \mathbf{1}_{[0,\frac{1}{2}s]} (B) + \tilde{A}(k) \right)^{-1}$$
par définition des opérateurs $\mathcal{T}_{V}(z(k))$ et $K(k,s)$. Par l'identité (\ref{eq2,131}), nous déduisons pour $\textup{Im} \hspace{0.5mm} z > 0$ l'inversibilité de $I + \mathcal{T}_{V}(z)$ avec pour inverse $$\left( I + \mathcal{T}_{V}(z) \right)^{-1} = I - J \vert V \vert^{1/2} \left( H - z \right)^{-1} \vert V \vert^{1/2}.$$
Or $\textup{Im} \hspace{0.5mm} z = \textup{Im} \hspace{0.5mm} k^{2}$. D'où l'inversibilité de l'opérateur $I + K(k,s)$ pour $\textup{Im} \hspace{0.5mm} k^{2} > \varsigma > 0$, $0 < s < \vert k \vert < s_{0}$ avec
\begin{align*}
\left\Vert (I + K(k,s))^{-1} \right\Vert & = O \left( 1 + \left\Vert \vert V \vert^{1/2} \left( H - z(k) \right)^{-1} \vert V \vert^{1/2} \right\Vert \right) = O \left( 1 + \vert \textup{Im} \hspace{0.5mm} k^{2} \vert^{-1} \right) \\
& = O \left( \varsigma^{-1} \right).
\end{align*}
Ce qui achève la preuve.
\end{preuve}

La Proposition \ref{prop2,16} ci-dessus montre pour $0 < s < \vert k \vert < s_{0}$ que
\begin{equation}\label{eq2,30}
\begin{aligned}
D(k,s) & = \prod_{j=1}^{O (n_{+}(s,pWp) + 1)} \big{(} 1 + \lambda_{j}(k,s) \big{)}\\
& = O \hspace{0.5mm} \textup{exp} \hspace{0.5mm} \big{(} O (n_{+}(s,pWp) + 1) \vert \textup{ln} \hspace{0.5mm} s \vert \big{)},
\end{aligned}
\end{equation}
où les $\lambda_{j}(k,s)$ sont les valeurs propres de $K := K(k,s)$ et satisfont $\vert \lambda_{j}(k,s) \vert = O \left( s^{-1} \right)$. De plus, puisque pour $\textup{Im} \hspace{0.5mm} k^{2} > \varsigma > 0$ et $0 < s < \vert k \vert < s_{0}$
$$D(k,s)^{-1} = \det \big{(} I + K \big{)}^{-1} = \det \big{(} I - K(I+K)^{-1} \big{)}.$$
Alors comme ci-dessus, pour $\textup{Im} \hspace{0.5mm} k^{2} > \varsigma > 0$
\begin{equation}\label{eq2,31}
\vert D(k,s) \vert \geq C \hspace{0.5mm} \textup{exp} \hspace{0.5mm} \big{(} - C (n_{+}(s,pWp) + 1) ( \vert \textup{ln} \hspace{0.5mm} \varsigma \vert + \vert \textup{ln} \hspace{0.5mm} s \vert ) \big{)}.
\end{equation}
Appliquons à présent le Lemme \ref{lem2,17} à la fonction $g(k) := D(rk,r)$ sur le domaine
$\Delta := \lbrace k \in \mathbb{C} : 1 < \vert k \vert < 2 \hspace{0.1cm} \textup{et} \hspace{0.1cm} \frac{\pi}{3} < \textup{Arg} \hspace{0.5mm} k < 2\pi + \frac{\pi}{6} \rbrace$ avec $\textup{Im} \hspace{0.5mm} k_{0}^{2} > \varsigma > 0$. Regarder les zéros de la fonction $g(k)$ sur un sous-domaine $\Delta' \Subset \Delta$, avec $0 < r < r \vert k \vert < r_{0} < \epsilon$, est équivalent par la Proposition \ref{prop2,16} à regarder les résonances $z(k)$ de l'opérateur $H$. Ainsi \eqref{eq2,181} s'obtient en combinant (\ref{eq2,30}) et (\ref{eq2,31}), ce qui conclut la preuve du Théorème \ref{theo2,18}.

\subsection{Preuve du Théorème \ref{theo2,24}}

Les notations sont celles de la sous-section 3.2. Avec la terminologie de valeur caractéristique, nous pouvons reformuler la Proposition \ref{prop2,10} comme suit:

\begin{prop}\label{prop2,23} 
Pour $k$ proche de $0$, les assertions suivantes sont équivalentes.

$\textup{(i)}$ $z_{1} := z(k_{1}) = k_{1}^{2}$ est une résonance de $H$ près de $0$,

$\textup{(ii)}$ $k_{1}$ est une valeur caractéristique de $I + \mathcal{T}_{V}(z(k))$. 
\\
De plus, la multiplicité de la résonance $z_{1}$ est égale à $\textup{Ind}_{\gamma} \hspace{0.5mm} \left( I + \mathcal{T}_{V}(\cdot) \right)$ la multiplicité de la valeur caractéristique $k_{1}$, $\gamma$ étant un cercle assez petit orienté positivement et contenant $k_{1}$ comme unique valeur caractéristique.
\end{prop}

Ainsi, l'étude des résonances $z(k) = k^{2}$ de $H_{e}$ près de $0$ peut se ramener à celle des valeurs caractéristiques de $$I + \mathcal{T}_{eV}(z(k)) = I - e \frac{T(ik)}{ik}$$ où $T(ik) := (\textup{sign} \hspace{0.5mm} V) B - ik A(k)$. Les opérateurs compacts $B$ et $A(k)$ sont définis dans la Proposition \ref{prop2,12}. En particulier pour $V > 0$, nous avons $T(0) = B$. Comme dit dans la Remarque \ref{rem2,1}, $(\textup{sign} \hspace{0.5mm} V) \tiny{\begin{pmatrix}
1 & 0\\
0 & 0
\end{pmatrix}} = \tiny{\begin{pmatrix}
1 & 0\\
0 & 0
\end{pmatrix}}$ suffit pour que $T(0) = B$. Puisque $T'(0) \Pi_{0}$ est compact, alors il existe une suite $(e_{n})_{n}$ qui tend vers l'infini telle que $I - e T'(0) \Pi_{0}$ soit inversible dès que $e \in \mathbb{R} \setminus \lbrace e_{n}, n \in \mathbb{N} \rbrace$. Plus précisément, nous pouvons choisir pour $e_{n}$ les inverses des valeurs propres de l'opérateur $T'(0) \Pi_{0}$.

Tournons nous vers la preuve du Théorème \ref{theo2,24}. Le point $\textup{(i)}$ s'obtient par une application du Lemme \ref{lem2,19} avec $z = ik/e$.

Par ailleurs, $\textup{(i)}$ montre que pour $\vert k \vert$ suffisamment petit, les résonances $z(k)$ de $H_{e}$ s'accumulent dans le secteur $\lbrace k \in D(0,\epsilon)^{\ast} : ik/e \in \mathcal{C}_{\alpha}(r,r_{0}) \rbrace$ pour tout $\alpha > 0$, où $\mathcal{C}_{\alpha}(r,r_{0})$ est défini par \eqref{eq2,33}. D'où en particulier lorsque $r \searrow 0$,
\begin{align*}
\# \lbrace z & = z(k) \in \textup{Res}(H_{e}) : r < \vert k \vert \leq r_{0} \rbrace \\
& = \# \lbrace z = z(k) \in \textup{Res}(H_{e}) : ik/e \in \mathcal{C}_{\alpha}(r,r_{0}) \rbrace + O(1).
\end{align*}
 Or nous avons $$\# \lbrace z = z(k) \in \textup{Res}(H_{e}) : ik/e \in \mathcal{C}_{\alpha}(r,r_{0}) \rbrace = \mathcal{N}(\mathcal{C}_{\alpha}(r,r_{0})).$$ De plus via (\ref{eq2,23}), $n(]r,r_{0}]) = n_{+} \left( r,\tiny{\frac{1}{2}}pWp \right) + O(1)$. Ainsi, le point $\textup{(ii)}$ découle directement du Lemme \ref{lem2,20} et des Lemmes \ref{lem2,21}, \ref{lem2,13}, \ref{lem2,14} et \ref{lem2,15}. Ceci conclut la preuve du Théorème \ref{theo2,24}.

\section{Résonances près de $\pm m$ de l'opérateur de Dirac}

Dans toute cette section, nous supposons que $V$ vérifie \eqref{eq1,14} (pour $n = 4$). La démarche reste similaire à celle de la section précédente. Nous traitons uniquement l'étude des résonances près de $m$. L'étude près de $-m$ se traite de la même manière (voir Remarque \ref{rem5,1}).

\subsection{Résonances}

Soit $D_{0}$ l'opérateur de Dirac défini par \eqref{eq1,11}. Nous définissons dans cette partie les résonances de l'opérateur $D := D_{0} + V$.

Soient $\textup{P}_{d}$ ($d$ pour Dirac) la projection orthogonale sur l'union des sous-espaces propres associés aux valeurs propres $\pm m$ de $D_{0}$ et $\textup{Q}_{d} := I - \textup{P}_{d}$. Par (\ref{eq1,12}),
\begin{equation}\label{eq3,7}
\textup{P}_{d} := \begin{pmatrix} 
   \textup{P} & 0 \\ 
   0 & \textup{P} 
\end{pmatrix}, \hspace{1cm} \textup{Q}_{d} := \textup{I} - \textup{P} = \begin{pmatrix} 
   \textup{Q} & 0 \\
   0 & \textup{Q} 
\end{pmatrix},
\end{equation}
où $\textup{P}$ et $\textup{Q}$ sont les projecteurs définis par (\ref{eq2,4}).
\\
Soit $z \in \mathbb{C} \setminus \sigma (D_{0})$. Nous avons
\begin{equation}\label{eq3,9}
\left( D_{0} - z \right)^{-1} = \left( D_{0} - z \right)^{-1} \textup{P}_{d} + \left( D_{0} - z \right)^{-1} \textup{Q}_{d}
\end{equation}
et $\left( D_{0} - z \right)^{-1} \textup{P}_{d} = (D_{0} + z) \left( D_{0}^{2} - z^{2} \right)^{-1} \textup{P}_{d}$. Par conséquent par (\ref{eq1,12}) et (\ref{eq1,13}), nous avons
\begin{equation}\label{eq3,10}
\begin{split}
\left( D_{0} - z \right)^{-1} & \textup{P}_{d}
 = \left[ p \otimes \mathcal{R}(z^{2} - m^{2}) \right] \left( \begin{smallmatrix}
   z + m & 0 & 0 & 0\\
   0 & 0 & 0 & 0\\
   0 & 0 & z - m & 0\\
   0 & 0 & 0 & 0
\end{smallmatrix} \right) \\
& + \left[ p \otimes \Pi_{3} \mathcal{R}(z^{2} - m^{2}) \right] \left( \begin{smallmatrix}
   0 & 0 & 1 & 0\\
   0 & 0 & 0 & 0\\
   1 & 0 & 0 & 0\\
   0 & 0 & 0 & 0
\end{smallmatrix} \right).
\end{split}
\end{equation}
Soit $z \in \mathbb{C}^{+}$. Pour $k \in \mathbb{C}_{1/2}^{+}$ défini par \eqref{eq2,10}, posons
\begin{equation}\label{eq3,11}
\frac{z-m}{z+m} = k^{2} \Longleftrightarrow z = z_{m}(k) := \frac{m(1+k^{2})}{1-k^{2}} \in \mathbb{C}^{+}.
\end{equation}

La première étape consiste à prolonger holomorphiquement $\left( D_{0} - z_{m}(k) \right)^{-1} \textup{P}_{d}$ au voisinage de $k = 0$.

\begin{lem}\label{prop3,22} 
Soient $\delta$ et $D(0,\eta)^{\ast}$ définis respectivement par \eqref{eq1,14} et \eqref{eq2,1100}.

$\textup{(i)}$ L'application $$k \longmapsto \left( \left( D_{0} - z_{m}(k) \right)^{-1} \textup{P}_{d} : \textup{e}^{-\delta \langle x_{3} \rangle} L^{2}(\mathbb{R}^{3}) \longrightarrow \textup{e}^{\delta \langle x_{3} \rangle} L^{2}(\mathbb{R}^{3}) \right)$$ admet un prolongement holomorphe de $\mathbb{C}_{1/2}^{+} \cap D(0,\eta)^{\ast}$ à $D(0,\eta)^{\ast}$ avec $\eta < \frac{\delta}{4m}$.

$\textup{(ii)}$ Soit $v_{12}(x_{12}) := \langle x_{12} \rangle^{-\alpha}$, $\alpha > 0$. L'application $$T_{v_{12}} : k \longmapsto v_{12}(x_{12}) \textup{e}^{-\delta \langle x_{3} \rangle} \left( D_{0} - z_{m}(k) \right)^{-1} \textup{P}_{d} \textup{e}^{-\delta \langle x_{3} \rangle}$$ se prolonge aussi holomorphiquement sur $D(0,\eta)^{\ast}$, à valeurs dans $S_{\infty} \left( L^{2}(\mathbb{R}^{3}) \right)$.
\end{lem}

\hspace{-0.54cm} \begin{preuve}
$\textup{(i)}$ Vus comme opérateurs de $\textup{e}^{-\delta \langle x_{3} \rangle} L^{2}(\mathbb{R}^{3})$ dans $\textup{e}^{\delta \langle x_{3} \rangle} L^{2}(\mathbb{R}^{3})$, soient $$L_{1}(k) := \left[ p \otimes \mathcal{R} \left( k^{2} (z+m)^{2} \right) \right] \left( \begin{smallmatrix}
   z + m & 0 & 0 & 0\\
   0 & 0 & 0 & 0\\
   0 & 0 & z - m & 0\\
   0 & 0 & 0 & 0
\end{smallmatrix} \right)$$ et $$L_{2}(k) := \left[ p \otimes \Pi_{3} \mathcal{R} \left( k^{2} (z+m)^{2} \right) \right] \left( \begin{smallmatrix}
   0 & 0 & 1 & 0\\
   0 & 0 & 0 & 0\\
   1 & 0 & 0 & 0\\
   0 & 0 & 0 & 0
\end{smallmatrix} \right).$$ Le noyau intégral de l'opérateur $N_{1}(k) := \textup{e}^{-\delta \langle x_{3} \rangle} \mathcal{R} \left( k^{2} (z+m)^{2} \right) \textup{e}^{-\delta \langle x_{3} \rangle}$ est 
\begin{equation}\label{eq3,111}
\textup{e}^{-\delta \langle x_{3} \rangle} \frac{i \textup{e}^{i k(z+m) \vert x_{3} -x'_{3} \vert}}{2 k(z+m)} \textup{e}^{-\delta \langle x'_{3} \rangle}.
\end{equation}
Il est dans $L^{2}(\mathbb{R}^{2})$ pour $\textup{Im} \hspace{0.5mm} k(z+m) = \textup{Im} \hspace{0.5mm} \frac{2mk}{1-k^{2}} > - \delta$. Donc si $\eta < \frac{\delta}{4m}$, on peut prolonger holomorphiquement $k \longmapsto L_{1}(k) \in \mathcal{L} \left( \textup{e}^{-\delta \langle x_{3} \rangle} L^{2}(\mathbb{R}^{3}),\textup{e}^{\delta \langle x_{3} \rangle} L^{2}(\mathbb{R}^{3}) \right)$ de $\mathbb{C}_{1/2}^{+} \cap D(0,\eta)^{\ast}$ à $D(0,\eta)^{\ast}$. De même se prolonge holomorphiquement l'application $k \longmapsto L_{2}(k) \in \mathcal{L} \left( \textup{e}^{-\delta \langle x_{3} \rangle} L^{2}(\mathbb{R}^{3}),\textup{e}^{\delta \langle x_{3} \rangle} L^{2}(\mathbb{R}^{3}) \right)$. D'où le prolongement holomorphe via (\ref{eq3,10}) de $k \longmapsto \left( D_{0} - z_{m}(k) \right)^{-1} \textup{P}_{d} \in \mathcal{L} \left( \textup{e}^{-\delta \langle x_{3} \rangle} L^{2}(\mathbb{R}^{3}),\textup{e}^{\delta \langle x_{3} \rangle} L^{2}(\mathbb{R}^{3}) \right)$ de\\ $\mathbb{C}_{1/2}^{+} \cap D(0,\eta)^{\ast}$ à $D(0,\eta)^{\ast}$ pour $\eta < \delta / 4m$.

$\textup{(ii)}$ Par (\ref{eq3,10})
\begin{equation}\label{eq3,15}
T_{v_{12}}(k) = \left[ v_{12}p \otimes N_{1}(k) \right] \left( \begin{smallmatrix}
   z + m & 0 & 0 & 0\\
   0 & 0 & 0 & 0\\
   0 & 0 & z - m & 0\\
   0 & 0 & 0 & 0
\end{smallmatrix} \right) + \left[ v_{12}p \otimes N_{2}(k) \right] \left( \begin{smallmatrix}
   0 & 0 & 1 & 0\\
   0 & 0 & 0 & 0\\
   1 & 0 & 0 & 0\\
   0 & 0 & 0 & 0
\end{smallmatrix} \right).
\end{equation}
On conclut en raisonnant comme dans la preuve du Lemme \ref{prop2,5}.
\end{preuve}
La deuxième étape consiste à prolonger holomorphiquement $\left( D_{0} - z \right)^{-1} \textup{Q}_{d}$ en la variable $z$. 

\begin{par}
\begin{lem}\label{prop3,23} 
Soient $\delta$ et $\zeta$ définis respectivement par \eqref{eq1,14} et \eqref{eq0}.

$\textup{(i)}$ L'application $$z \longmapsto \left( \left( D_{0} - z \right)^{-1} \textup{Q}_{d} : \textup{e}^{-\delta \langle x_{3} \rangle} L^{2}(\mathbb{R}^{3}) \longrightarrow \textup{e}^{\delta \langle x_{3} \rangle} L^{2}(\mathbb{R}^{3}) \right)$$ se prolonge holomorphiquement de $\mathbb{C}^{+}$ à $\mathbb{C} \setminus \left\lbrace \left( -\infty,-\sqrt{m^{2} + \zeta} \right] \cup \left[ \sqrt{m^{2} + \zeta},\infty \right) \right\rbrace$.

$\textup{(ii)}$ Soit $v_{12}(x_{12}) := \langle x_{12} \rangle^{-\alpha}$, $\alpha > 0$. L'application $$L_{v_{12}} : z \longmapsto v_{12}(x_{12}) \textup{e}^{-\delta \langle x_{3} \rangle} \left( D_{0} - z \right)^{-1} \textup{Q}_{d} \textup{e}^{-\delta \langle x_{3} \rangle}$$ se prolonge aussi holomorphiquement à $\mathbb{C} \setminus \left\lbrace \left( -\infty,-\sqrt{m^{2} + \zeta} \right] \cup \left[ \sqrt{m^{2} + \zeta},\infty \right) \right\rbrace$, à valeurs dans $S_{\infty} \left( L^{2}(\mathbb{R}^{3}) \right)$.
\end{lem}

\hspace{-0.54cm} \begin{preuve}
$\textup{(i)}$ Soit $z \in \mathbb{C}^{+}$. L'égalité $$\left( D_{0} - z \right)^{-1} = D_{0}^{-1} + z \left( 1 + z D_{0}^{-1} \right) \left( D_{0}^{2} - z^{2} \right)^{-1}$$ donne
\begin{align*}
\textup{e}^{-\delta \langle x_{3} \rangle} \left( D_{0} - z \right)^{-1} \textup{Q}_{d} \textup{e}^{-\delta \langle x_{3} \rangle} & = \textup{e}^{-\delta \langle x_{3} \rangle} D_{0}^{-1} \textup{Q}_{d} \textup{e}^{-\delta \langle x_{3} \rangle} \\ 
& + z \textup{e}^{-\delta \langle x_{3} \rangle} \left( 1 + z D_{0}^{-1} \right) \left( D_{0}^{2} - z^{2} \right)^{-1} \textup{Q}_{d} \textup{e}^{-\delta \langle x_{3} \rangle}.
\end{align*}
L'identité (\ref{eq1,13}) et \eqref{eq3,7} entraînent que $\left( D_{0}^{2} - z^{2} \right)^{-1} \textup{Q}_{d}$ est donné par 
$$\left( \begin{smallmatrix}
\left( H_{0}^{-} + m^{2} - z^{2} \right)^{-1} Q & 0 & 0 & 0\\
   0 & \left( H_{0}^{+} + m^{2} - z^{2} \right)^{-1} & 0 & 0\\
   0 & 0 & \left( H_{0}^{-} + m^{2} - z^{2} \right)^{-1} Q & 0\\
   0 & 0 & 0 & \left( H_{0}^{+} + m^{2} - z^{2} \right)^{-1}
\end{smallmatrix} \right).$$ Ainsi, $\mathbb{C} \setminus \left\lbrace \left( -\infty,-\sqrt{m^{2} + \zeta} \right] \cup \left[ \sqrt{m^{2} + \zeta},\infty \right) \right\rbrace \ni z \longrightarrow \left( D_{0}^{2} - z^{2} \right)^{-1} \textup{Q}_{d}$ est bien définie et analytique car $\mathbb{C} \setminus [\zeta,\infty)$ est contenu dans l'ensemble résolvent de l'opérateur $H_{0}^{-}$ restreint à $Q D(H_{0}^{-})$, et de l'opérateur $H_{0}^{+}$ restreint à $D(H_{0}^{+})$. L'opérateur $\textup{e}^{-\delta \langle x_{3} \rangle} \left( D_{0} - z \right)^{-1} \textup{Q}_{d} \textup{e}^{-\delta \langle x_{3} \rangle}$ se prolonge alors holomorphiquement à l'ensemble des points $z \in \mathbb{C} \setminus \left\lbrace \left( -\infty,-\sqrt{m^{2} + \zeta} \right] \cup \left[ \sqrt{m^{2} + \zeta},\infty \right) \right\rbrace$.

$\textup{(ii)}$ Par la décomposition ci-dessus, $$L_{v_{12}}(z) = v_{12} \textup{e}^{-\delta \langle x_{3} \rangle} D_{0}^{-1} \textup{Q}_{d} \textup{e}^{-\delta \langle x_{3} \rangle} + z v_{12} \textup{e}^{-\delta \langle x_{3} \rangle} \left( 1 + z D_{0}^{-1} \right) \left( D_{0}^{2} - z^{2} \right)^{-1} \textup{Q}_{d} \textup{e}^{-\delta \langle x_{3} \rangle}.$$ Un raisonnement similaire à celui de la preuve de la Proposition 5.3 de \cite{tda} permet de montrer que $L_{v_{12}}(z)$ est dans une certaine classe $S_{q}$, donc compact. On conclut la preuve comme pour le Lemme \ref{prop2,6}.
\end{preuve}

Le lemme suivant découle des Lemmes \ref{prop3,22} et \ref{prop3,23}.

\begin{lem}\label{lem3,26} 
Soit $D(0,\eta)^{\ast}$ défini par \eqref{eq2,1100}. Supposons que $V$ vérifie \eqref{eq1,14} pour $n = 4$. L'application $$\mathbb{C}_{1/2}^{+} \cap D(0,\eta)^{\ast} \ni k \longmapsto \mathcal{M}_{V}(z_{m}(k)) := J \vert V \vert^{1/2} \left( D_{0} - z_{m}(k) \right)^{-1} \vert V \vert^{1/2},$$ avec $J := \textup{sign} \hspace{0.5mm} V$, se prolonge analytiquement à $D(0,\eta)^{\ast}$ à valeurs dans $S_{\infty} \left( L^{2}(\mathbb{R}^{3}) \right)$. Ce prolongement est encore noté $\mathcal{M}_{V}(z_{m}(k))$.
\end{lem}

\hspace{-0.54cm} L'identité $$\left( D - z \right)^{-1} \left( 1 + V(D_{0} - z)^{-1} \right) = (D_{0} - z)^{-1},$$ donne 
\begin{align*}
\textup{e}^{-\delta \langle x_{3} \rangle} \left( D - z \right)^{-1} \textup{e}^{-\delta \langle x_{3} \rangle} & = \textup{e}^{-\delta \langle x_{3} \rangle} (D_{0} - z)^{-1} \textup{e}^{-\delta \langle x_{3} \rangle} \\ 
& \times \left( 1 + \textup{e}^{\delta \langle x_{3} \rangle} V(D_{0} - z)^{-1} \textup{e}^{-\delta \langle x_{3} \rangle} \right)^{-1}.
\end{align*}
Par le Lemme \ref{lem3,26}, la fonction $k \longmapsto \textup{e}^{\delta \langle x_{3} \rangle}V(D_{0} - z_{m}(k))^{-1} \textup{e}^{-\delta \langle x_{3} \rangle}$ est holomorphe à valeur opérateur compact dans $L^{2}(\mathbb{R}^{3})$ inversible en au moins un point. D'où par le théorème analytique de Fredholm, $$k \longmapsto \left( 1 + \textup{e}^{\delta \langle x_{3} \rangle} V(D_{0} - z_{m}(k))^{-1} \textup{e}^{-\delta \langle x_{3} \rangle} \right)^{-1}$$ admet un prolongement méromorphe de $\mathbb{C}_{1/2}^{+} \cap D(0,\eta)^{\ast}$ à $D(0,\eta)^{\ast}$. Ce qui nous permet de définir les résonances de $D$ près de $m$.

\begin{prop}\label{prop3,25} 
Soient $\mathbb{C}_{1/2}^{+}$ défini par \eqref{eq2,10} et $D(0,\eta)^{\ast}$ par \eqref{eq2,1100}. L'application $$k \longmapsto \left( \left( D - z_{m}(k) \right)^{-1} : \textup{e}^{-\delta \langle x_{3} \rangle} L^{2}(\mathbb{R}^{3}) \longrightarrow \textup{e}^{\delta \langle x_{3} \rangle} L^{2}(\mathbb{R}^{3}) \right)$$ admet un prolongement méromorphe de $\mathbb{C}_{1/2}^{+} \cap D(0,\eta)^{\ast}$ à $D(0,\eta)^{\ast}$. Ce prolongement est encore noté $R(z_{m}(k))$.
\end{prop}

De manière analogue à la Définition \ref{def2,11}, nous avons la 

\begin{déf}\label{def3,28} 
\textup{Nous définissons les résonances de l'opérateur $D$ (près de $m$) comme étant les pôles du prolongement méromorphe noté $R(z)$, de la résolvante $\left( D - z \right)^{-1}$ dans $\mathcal{L} \left( \textup{e}^{-\delta \langle x_{3} \rangle} L^{2}(\mathbb{R}^{3}),\textup{e}^{\delta \langle x_{3} \rangle} L^{2}(\mathbb{R}^{3}) \right)$. La multiplicité d'une résonance $z_{0} := z_{m}(k_{0})$ est définie par 
\begin{equation}\label{eq3,16}
\textup{mult}(z_{0}) := \textup{Ind}_{\gamma} \hspace{0.5mm} \left( I + \mathcal{M}_{V}(z_{m}(\cdot)) \right),
\end{equation}
où $\gamma$ est un cercle assez petit orienté positivement contenant $k_{0}$ comme unique point vérifiant que $z_{m}(k_{0})$ est résonance de $D$, et $\mathcal{M}_{V}(z_{m}(\cdot))$ est défini au Lemme \ref{lem3,26}.}
\end{déf}

La proposition suivante est l'analogue dans le cas Dirac de la Proposition \ref{prop2,10}.

\begin{prop}\label{prop3,27} 
Pour $k$ proche de $0$, les assertions suivantes sont équivalentes:

$\textup{(i)}$ $z_{m}(k)$ est un pôle de $R(z_{m}(k))$,

$\textup{(ii)}$ $-1$ est une valeur propre de $\mathcal{M}_{V}(z_{m}(k)) := J \vert V \vert^{1/2} R_{0}(z_{m}(k)) \vert V \vert^{1/2}$.
\end{prop}
\end{par}

\subsection{Preuve du Théorème \ref{theo3,31}}

\begin{par}
Posons $z = z_{m}(k)$ et considérons $\mathcal{M}_{V}(z_{m}(k))$ défini au Lemme \ref{lem3,26}. Rappelons que $p$ est la projection orthogonale sur $\textup{ker} \hspace{0.5mm} H_{12}^{-}$ défini par \eqref{eq1.161}. Via \eqref{eq3,9} et \eqref{eq3,10}, nous avons $\mathcal{M}_{V}(z_{m}(k)) = \mathcal{M}_{1}^{V}(k) + \mathcal{M}_{2}^{V}(k)$ où 
$$\mathcal{M}_{1}^{V}(k) := J \vert V \vert^{1/2} \left[ p \otimes \mathcal{R} \left( k^{2} (z+m)^{2} \right) \right] \left( \begin{smallmatrix}
   z + m & 0 & 0 & 0\\
   0 & 0 & 0 & 0\\
   0 & 0 & z - m & 0\\
   0 & 0 & 0 & 0
\end{smallmatrix} \right) \vert V \vert^{1/2},$$
\begin{align*}
\mathcal{M}_{2}^{V}(k) :=  J \vert & V \vert^{1/2} \left[ p \otimes \Pi_{3} \mathcal{R} \left( k^{2} (z+m)^{2} \right) \right] \left( \begin{smallmatrix}
   0 & 0 & 1 & 0\\
   0 & 0 & 0 & 0\\
   1 & 0 & 0 & 0\\
   0 & 0 & 0 & 0
\end{smallmatrix} \right) \vert V \vert^{1/2} \\ 
& + J \vert V \vert^{1/2} \left( H_{0} - z \right)^{-1} \textup{Q}_{d} \vert V \vert^{1/2}.
\end{align*}
L'opérateur $\mathcal{M}_{2}^{V}(k)$ est holomorphe près de $0$ à valeurs dans $S_{\infty} \left( L^{2}(\mathbb{R}^{3}) \right)$. Considérons à présent $\mathcal{M}_{1}^{V}(k)$ pour $k \in D(0,\eta)^{\ast}$ défini par \eqref{eq2,1100}. Le noyau intégral de l'opérateur $N_{1}(k) := \textup{e}^{-\delta \langle x_{3} \rangle} \mathcal{R} \left( k^{2} (z+m)^{2} \right) \textup{e}^{-\delta \langle x_{3} \rangle}$ est donné par \eqref{eq3,111}. D'où la décomposition
\begin{equation}\label{eq3,17}
N_{1}(k) = \frac{1}{k(z+m)}t_{1} + b_{1}(k),
\end{equation}
où $t_{1}$ est défini par \eqref{eq2,16}
et $b_{1}(k)$ l'opérateur de Hilbert-Schmidt (sur $D(0,\eta)^{\ast}$) ayant pour noyau intégral 
\begin{equation}\label{eq3,19}
\textup{e}^{-\delta \langle x_{3} \rangle} i \frac{ \textup{e}^{ i k(z+m) \vert x_{3} -x'_{3} \vert} - 1}{2 k(z+m)} \textup{e}^{-\delta \langle x'_{3} \rangle}.
\end{equation}
Par conséquent
\begin{equation}\label{eq3,20}
\begin{aligned}
\mathcal{M}_{1}^{V}(k) = \frac{iJ}{k(z+m)} \times \frac{1}{2} \vert V \vert^{1/2} & \left[ p \otimes \tau_{1} \right] \left( \begin{smallmatrix}
   z + m & 0 & 0 & 0\\
   0 & 0 & 0 & 0\\
   0 & 0 & z - m & 0\\
   0 & 0 & 0 & 0
\end{smallmatrix} \right) \vert V \vert^{1/2} \\
& + J \vert V \vert^{1/2} \left[ p \otimes c_{1}(k) \right] \left( \begin{smallmatrix}
   z + m & 0 & 0 & 0\\
   0 & 0 & 0 & 0\\
   0 & 0 & z - m & 0\\
   0 & 0 & 0 & 0
\end{smallmatrix} \right) \vert V \vert^{1/2},
\end{aligned}
\end{equation}
où $\tau_{1}$ et $c_{1}(k)$ sont des opérateurs dans $L^{2}(\mathbb{R})$ ayant respectivement pour noyau intégral $1$ et
\begin{equation}\label{eq3,21}
\frac{ 1 - \textup{e}^{ i k(z+m) \vert x_{3} -x'_{3} \vert}}{2 i k(z+m)}.
\end{equation}
Soient les opérateurs bornés $K_{\pm} : L^{2}(\mathbb{R}^{3},\mathbb{C}^{4}) \rightarrow L^{2}(\mathbb{R}^{2},\mathbb{C}^{4})$ définis par 
\begin{equation}\label{eq3,210}
\begin{split}
(K_{+}\psi)(x_{12}) & := \int_{\mathbb{R}^{3}} \mathcal{P}(x_{12},x'_{12})\left( \begin{smallmatrix}
   1 & 0 & 0 & 0\\
   0 & 0 & 0 & 0\\
   0 & 0 & 0 & 0\\
   0 & 0 & 0 & 0
\end{smallmatrix} \right) \vert V \vert^{1/2}(x'_{12},x'_{3})\psi(x'_{12},x'_{3}) dx'_{12}dx'_{3},\\
(K_{-}\psi)(x_{12}) & := \int_{\mathbb{R}^{3}} \mathcal{P}(x_{12},x'_{12})\left( \begin{smallmatrix}
   0 & 0 & 0 & 0\\
   0 & 0 & 0 & 0\\
   0 & 0 & 1 & 0\\
   0 & 0 & 0 & 0
\end{smallmatrix} \right) \vert V \vert^{1/2}(x'_{12},x'_{3})\psi(x'_{12},x'_{3}) dx'_{12}dx'_{3},
\end{split}
\end{equation}
où $\mathcal{P}(\cdot,\cdot)$ est le noyau intégral de la projection $p$. Les opérateurs adjoints\\ $K^{\ast}_{\pm} : L^{2}(\mathbb{R}^{2},\mathbb{C}^{4}) \rightarrow L^{2}(\mathbb{R}^{3},\mathbb{C}^{4})$ sont donnés par 
\begin{align*}
(K^{\ast}_{+}\varphi)(x_{12},x_{3}) & = \vert V \vert^{1/2}(x_{12},x_{3}) \left( \begin{smallmatrix}
   1 & 0 & 0 & 0\\
   0 & 0 & 0 & 0\\
   0 & 0 & 0 & 0\\
   0 & 0 & 0 & 0
\end{smallmatrix} \right) (p\varphi)(x_{12}),\\
(K^{\ast}_{-}\varphi)(x_{12},x_{3}) & = \vert V \vert^{1/2}(x_{12},x_{3}) \left( \begin{smallmatrix}
   0 & 0 & 0 & 0\\
   0 & 0 & 0 & 0\\
   0 & 0 & 1 & 0\\
   0 & 0 & 0 & 0
\end{smallmatrix} \right) (p\varphi)(x_{12}),
\end{align*}
et nous avons
\begin{align*}
K^{\ast}_{+} K_{+} & = \vert V \vert^{1/2} [p \otimes \tau_{1}] \left( \begin{smallmatrix}
   1 & 0 & 0 & 0\\
   0 & 0 & 0 & 0\\
   0 & 0 & 0 & 0\\
   0 & 0 & 0 & 0
\end{smallmatrix} \right) \vert V \vert^{1/2},\\
K^{\ast}_{-} K_{-} & = \vert V \vert^{1/2} [p \otimes \tau_{1}] \left( \begin{smallmatrix}
   0 & 0 & 0 & 0\\
   0 & 0 & 0 & 0\\
   0 & 0 & 1 & 0\\
   0 & 0 & 0 & 0
\end{smallmatrix} \right) \vert V \vert^{1/2}.
\end{align*}
Finalement (\ref{eq3,20}) devient
$$\mathcal{M}_{1}^{V}(k) = \frac{iJ}{2k} K^{\ast}_{+} K_{+} + \frac{iJk}{2} K^{\ast}_{-} K_{-} + J \vert V \vert^{1/2} \left[ p \otimes c_{1}(k) \right] \left( \begin{smallmatrix}
   z + m & 0 & 0 & 0\\
   0 & 0 & 0 & 0\\
   0 & 0 & z - m & 0\\
   0 & 0 & 0 & 0
\end{smallmatrix} \right) \vert V \vert^{1/2}.$$
D'où la

\begin{prop}\label{prop3,29} Pour $k \in D(0,\eta)^{\ast}$,
\begin{equation}\label{eq3,22}
\mathcal{M}_{V}(z_{m}(k)) = \frac{iJ}{k} B_{d} + A_{d}(k), \hspace{0.5cm} (d \hspace{0.94mm} pour \hspace{0.7mm} Dirac)
\end{equation}
où $J := \textup{sign} \hspace{0.5mm} V$, $B_{d}$ l'opérateur auto-adjoint positif défini par
\begin{equation}\label{eq3,23}
B_{d} := \frac{1}{2} K^{\ast}_{+} K_{+},
\end{equation} 
$K_{+}$ donné par \eqref{eq3,210}, et $A_{d}(k) \in S_{\infty} \left( L^{2}(\mathbb{R}^{3}) \right)$ holomorphe sur $D(0,\eta)^{\ast}$
$$A_{d}(k) := \frac{iJk}{2} K^{\ast}_{-} K_{-} + J \vert V \vert^{1/2} \left[ p \otimes c_{1}(k) \right] \left( \begin{smallmatrix}
   z + m & 0 & 0 & 0\\
   0 & 0 & 0 & 0\\
   0 & 0 & z - m & 0\\
   0 & 0 & 0 & 0
\end{smallmatrix} \right) \vert V \vert^{1/2} + \mathcal{M}_{2}^{V}(k).$$
\end{prop}

\begin{rem}\label{rem5,1}
\textup{Pour regarder les résonances près de $-m$, on effectue le changement de variables $\frac{z+m}{z-m} = k^{2}$ où $k \in \mathbb{C}_{1/2}^{-} := \lbrace k \in \mathbb{C}^{+} : \textup{Im} \hspace{0.5mm} k^{2} < 0 \rbrace$, de sorte que $z = z_{-m}(k) := \frac{-m(1+k^{2})}{1-k^{2}} \in \mathbb{C}^{+}$. De plus dans la Proposition \ref{prop3,29} ci-dessus, nous aurons pour $k \in D(0,\eta)^{\ast}$ défini par \eqref{eq2,1100} $$\mathcal{M}_{V}(z_{-m}(k)) = -\frac{iJ}{k} B_{d} + A_{d}(k),$$ où les opérateurs $B_{d}$ et $A_{d}(k)$ sont donnés par $B_{d} := \frac{1}{2} K^{\ast}_{-} K_{-}$, $$A_{d}(k) := -\frac{iJk}{2} K^{\ast}_{+} K_{+} - J \vert V \vert^{1/2} \left[ p \otimes c_{1}(k) \right] \left( \begin{smallmatrix}
   z + m & 0 & 0 & 0\\
   0 & 0 & 0 & 0\\
   0 & 0 & z - m & 0\\
   0 & 0 & 0 & 0
\end{smallmatrix} \right) \vert V \vert^{1/2} + \mathcal{M}_{2}^{V}(k),$$ avec 
\vspace*{-0.3cm}
\begin{align*}
\mathcal{M}_{2}^{V}(k) :=  J \vert & V \vert^{1/2} \left[ p \otimes \Pi_{3} \mathcal{R} \left( k^{2} (z-m)^{2} \right) \right] \left( \begin{smallmatrix}
   0 & 0 & 1 & 0\\
   0 & 0 & 0 & 0\\
   1 & 0 & 0 & 0\\
   0 & 0 & 0 & 0
\end{smallmatrix} \right) \vert V \vert^{1/2} \\ 
& + J \vert V \vert^{1/2} \left( H_{0} - z \right)^{-1} \textup{Q}_{d} \vert V \vert^{1/2}.
\end{align*}}
\end{rem}

Soient $W_{\pm}$ les fonctions définies par \eqref{eq2,22}. Nous avons
$$K_{+}K^{\ast}_{+} = \left( \begin{smallmatrix}
   1 & 0 & 0 & 0\\
   0 & 0 & 0 & 0\\
   0 & 0 & 0 & 0\\
   0 & 0 & 0 & 0
\end{smallmatrix} \right) p W_{+} p \hspace{0.6cm} \textup{et} \hspace{0.6cm} 
K_{-}K^{\ast}_{-} =  \left( \begin{smallmatrix}
   0 & 0 & 0 & 0\\
   0 & 0 & 0 & 0\\
   0 & 0 & 1 & 0\\
   0 & 0 & 0 & 0
\end{smallmatrix} \right) p W_{-} p.$$
Ainsi pour $s > 0$,
\begin{align*}
n_{+} \big{(} s,\frac{1}{2} K^{\ast}_{+} K_{+} \big{)} & = n_{+} \big{(} 2s,K^{\ast}_{+} K_{+} \big{)} = n_{+} \big{(} 2s,K_{+}K^{\ast}_{+} \big{)}\\
& = n_{+} \left( 2s,\left( \begin{smallmatrix}
   1 & 0 & 0 & 0\\
   0 & 0 & 0 & 0\\
   0 & 0 & 0 & 0\\
   0 & 0 & 0 & 0
\end{smallmatrix} \right) p W_{+} p \right)\\
& = n_{+} \left( 2s,p W_{+} p \right).
\end{align*}
C'est-à-dire
\begin{equation}\label{eq3,25}
n_{+} \left( s,B_{d} \right) = n_{+} \left( 2s,p W_{+} p \right).
\end{equation}

\begin{prop}\label{prop3,30} 
Soit $s_{0} < \eta$ assez petit. Pour $0 < s < \vert k \vert < s_{0}$, 

$\textup{(i)}$ $z_{m}(k)$ est une résonance de $D$ (près de m) si et seulement si $k$ est un zéro de 
\begin{equation}\label{eq3,26}
D_{d}(k,s) := \det \big{(} I + K_{d}(k,s) \big{)},
\end{equation}
où $K_{d}(k,s)$ est un opérateur de rang fini analytique en $k$, vérifiant  
$$\textup{rang} \hspace{0.6mm} K_{d}(k,s) = O \big{(} n_{+}(s,pW_{+}p) + 1 \big{)}, \hspace{0.3cm} \left\Vert K_{d}(k,s) \right\Vert = O \left( s^{-1} \right),$$
uniformément pour $s < \vert k \vert < s_{0}$.

$\textup{(ii)}$ Si $z_{0} := z_{m}(k_{0})$ est une résonance de $D$, alors 
\begin{equation}\label{eq3,27}
\textup{mult}(z_{0}) = \textup{Ind}_{\gamma} \hspace{0.5mm} \left( I + K_{d}(k,s) \right) = \textup{m}(k_{0}),
\end{equation}
$\gamma$ étant choisi comme dans la \textup{Définition \ref{def3,28}} et $\textup{m}(k_{0})$ désignant la multiplicité de $k_{0}$ en tant que zéro du déterminant $D_{d}(k,s)$.

$\textup{(iii)}$ Si $\textup{Im} \hspace{0.5mm} k^{2} > \varsigma > 0$, l'opérateur $I + K_{d}(k,s)$ est inversible avec 
$$\left\Vert \big{(} I + K_{d}(k,s) \big{)}^{-1} \right\Vert = O \left( \varsigma^{-1} \right),$$
uniformément pour $s < \vert k \vert < s_{0}$.
\end{prop}

\hspace{-0.54cm} On conclut alors la preuve du Théorème \ref{theo3,31} comme pour le Théorème \ref{theo2,18} en remplaçant le determinant $D(k,s)$ par $D_{d}(k,s)$, et l'opérateur $K(k,s)$ par $K_{d}(k,s)$.
\end{par}

\subsection{Preuve du Théorème \ref{theo3,32}} 

La Proposition \ref{prop3,27} peut être reformulée de la manière suivante:

\begin{prop}\label{prop3,31} 
Pour $k$ près de $0$, les assertions suivantes sont équivalentes: 

$\textup{(i)}$ $z_{0} := z_{m}(k_{0})$ est une résonance de $D$ près de $m$,

$\textup{(ii)}$ $k_{0}$ est une valeur caractéristique de $I + \mathcal{M}_{V}(z_{m}(k))$.

\hspace{-0.54cm} De plus, la multiplicité de la résonance $z_{0}$ est égale à $\textup{Ind}_{\gamma} \hspace{0.5mm} \left( I + \mathcal{M}_{V}(\cdot) \right)$ la multiplicité de la valeur caractéristique $k_{0}$, $\gamma$ étant un cercle assez petit orienté positivement et contenant $k_{0}$ comme unique valeur caractéristique.
\end{prop}

\hspace{-0.54cm} L'étude des résonances $z_{m}(k)$ de $D_{\varepsilon}$ près de $m$ se ramène ainsi à celle des valeurs caractéristiques de l'opérateur $$I + \mathcal{M}_{\varepsilon V}(z_{0}(k)) = I - \varepsilon \frac{T_{d}(ik)}{ik}$$ où $T_{d}(ik) := (\textup{sign} \hspace{0.5mm} V) B_{d} - ik A_{d}(k)$. Les opérateurs compacts $B_{d}$ et $A_{d}(k)$ sont définis dans la Proposition \ref{prop3,29}. On conclut alors la preuve du Théorème \ref{theo3,32} comme pour le Théorème \ref{theo2,24}.

\section{Appendice}

Dans cette appendice, nous rappelons les notions d'indice (le long d'un contour fermé orienté positivement) d'une fonction holomorphe (scalaire) et d'une fonction méromorphe finie (voir p. e. la Définition 2.1 de \cite{bo}). 

\hspace{-0.54cm} Soit $f$ une fonction holomorphe sur un voisinage d'un contour fermé $\gamma$. L'indice de $f$ le long du contour $\gamma$ orienté positivement est défini par la quantité $$\textup{ind}_{\gamma} \hspace{0.5mm} f := \frac{1}{2i\pi} \int_{\gamma} \frac{f'(z)}{f(z)} dz.$$ Notons que si $f$ est holomorphe sur un domaine $\Omega$ tel que $\partial \Omega = \gamma$, alors par le Théorème des résidus $\textup{ind}_{\gamma} \hspace{0.5mm} f$ est le nombre de zéros de la fonction $f$ contenu dans $\Omega$, comptés avec leur multiplicité. Soient à présent un domaine ouvert borné $D \subseteq \mathbb{C}$, de frontière $C^{1}$ par morceaux, $Z \subset D$ un ensemble fini et $A : \overline{D} \backslash Z \longrightarrow \textup{GL}(E)$ une fonction méromorphe finie et de Fredholm aux points de $Z$. L'indice de l'opérateur $A$ sur le contour $\partial \Omega$ est défini par $$\textup{Ind}_{\partial \Omega} \hspace{0.5mm} A := \frac{1}{2i\pi} \textup{tr} \int_{\partial \Omega} A'(z)A(z)^{-1} dz = \frac{1}{2i\pi} \textup{tr} \int_{\partial \Omega} A(z)^{-1} A'(z) dz.$$ Nous avons les propriétés suivantes: $\textup{Ind}_{\partial \Omega} \hspace{0.5mm} A_{1} A_{2} = \textup{Ind}_{\partial \Omega} \hspace{0.5mm} A_{1} + \textup{Ind}_{\partial \Omega} \hspace{0.5mm} A_{2}$, et pour $K(z)$ un opérateur à trace, $\textup{Ind}_{\partial \Omega} \hspace{0.5mm} (I+K)= \textup{ind}_{\partial \Omega} \hspace{0.5mm} \det \hspace{0.5mm} (I + K)$. Pour plus de détails sur l'indice d'un opérateur sur un contour, nous renvoyons au Chapitre 4 de \cite{goh}.

\textit{Remerciements} Ce travail est soutenu par le programme ANR NOSEVOL-11-B501-019, NONAa.

\end{document}